\documentclass[smallcondensed]{svjour3} 
\usepackage[utf8]{inputenc}
\usepackage{amsmath}
\usepackage{amsfonts}
\usepackage{booktabs}
\usepackage{graphicx}
\usepackage{multirow}
\usepackage{natbib}
\usepackage{tikz}
\usepackage{comment}
\usepackage{wrapfig}
\usetikzlibrary{decorations.pathreplacing}

\smartqed

\if@abbrvbib
\else
\fi

\def\makeheadbox{\relax}

\bibpunct{(}{)}{;}{a}{,}{,}
\usepackage{algorithmic}
\usepackage{algorithm}

\newcommand\numberthis{\addtocounter{equation}{1}\tag{\theequation}}

\usepackage[title]{appendix}

\title{Mint: MDL-based approach for Mining INTeresting Numerical Pattern Sets}
\author{Tatiana Makhalova \and Sergei O. Kuznetsov \and Amedeo Napoli}
\institute{T. Makhalova\at
              Universit\'{e} de Lorraine, CNRS, Inria, LORIA, F-54000 Nancy, France\\
              \email{tatiana.makhalova@inria.fr}           
           \and
           S. O. Kuznetsov \at
               National Research University Higher School of Economics, Moscow, Russia\\
              \email{skuznetsov@hse.ru}     
            \and 
            A. Napoli \at
            Universit\'{e} de Lorraine, CNRS, Inria, LORIA, F-54000 Nancy, France\\
              \email{amedeo.napoli@loria.fr}   
}
\date{}

\begin{document}

\maketitle

\begin{abstract}

Pattern mining is well established in data mining research, especially for mining binary datasets.
Surprisingly, there is much less work about numerical pattern mining and this research area remains under-explored.
In this paper we propose \textsc{Mint}, an efficient MDL-based algorithm for mining numerical datasets.
The MDL principle is a robust and reliable framework widely used in pattern mining, and as well in subgroup discovery.
In \textsc{Mint} we reuse MDL for discovering useful patterns and returning a set of non-redundant overlapping patterns with well-defined boundaries and covering meaningful groups of objects.
\textsc{Mint} is not alone in the category of numerical pattern miners based on MDL.
In the experiments presented in the paper we show that \textsc{Mint} outperforms competitors among which \textsc{Slim} and \textsc{RealKrimp}.

\end{abstract}

%
\section{Introduction}

The objective of pattern mining is to discover a small set of interesting patterns that describe together a large portion of a dataset and can be easily interpreted and reused. 
Actually pattern mining encompasses a large variety of algorithms in knowledge discovery and data mining aimed at analyzing datasets \citep{aggarwal2014frequent5interesting}. 
Present approaches in pattern mining are aimed at discovering an interesting \textit{pattern set} rather than a set of individually interesting patterns, where the quality of patterns is evaluated w.r.t. both the dataset and other patterns. 
One common theoretical basis for pattern set mining relies on the Minimum Description Length principle (MDL)~\citep{grunwald2007minimum}, which is applied to many types of patterns, e.g. itemsets~\citep{vreeken2011krimp}, patterns of arbitrary shapes in 2-dimensional data~\citep{faas2020vouw}, sequences~\citep{hinrichs2017characterising}, graphs~\citep{bariatti2020graphmdl}, etc.

Contrasting the recent advances in pattern mining, algorithms for mining numerical data appear to be insufficiently explored. 
To date, one of the most common way to mine numerical pattern sets relies on the application of itemset mining to binarized datasets.
This is discussed below in more detail but before we would like to mention an alternative to numeric pattern mining which is ``clustering''.

In the last years, clustering algorithms have been extensively developed and many different and efficient approaches have been proposed \citep{Jain10,CraenendonckDB17,Jeantet0G20}.
However, there is an important conceptual difference between pattern mining and clustering. 
In pattern mining the \textit{description} comes first, while in clustering the primacy is given to the  \textit{object similarity}.
In other words, numerical pattern mining is more interested in the description of a group of objects in terms of a set of attributes related to these objects, while clustering focuses more on the detection of these groups of objects based on their commonalities as measured by a similarity or a distance.
The former entails some requirements for the ease of interpretation, i.e., the resulting patterns should describe a region in the ``attribute space'' that is easy to interpret.
By contrast, in clustering, the focus is put on groups of objects or instances. 
The clusters can be constrained to have certain shapes, e.g., spheres in \textsc{K-means} or \textsc{DBSCAN}, but still the similarity of objects remains the most important characteristic of clusters.
For example, clustering techniques such as agglomerative single-linkage clustering in a multidimensional space may return clusters of very complex shapes.
Usually no attention is paid to these shapes while this is one of the most important preoccupation in numerical pattern mining.

Accordingly, in this paper, we propose an MDL-based approach to numerical pattern set mining called \textsc{Mint} for ``Mining INTeresting Numerical Pattern Sets''. 
\textsc{Mint} computes numerical patterns as $m$-dimensional hyper-rectangles which are products of $m$ intervals, where the intervals are related to the attributes and their values.
The main benefits of the \textsc{Mint} approach are that
(i) \textsc{Mint} does not need to explore the pattern space in advance as candidate to become optimal patterns are computed on the fly,
(ii) the total number of explored patterns is at most cubic (and it is at most quadratic at each iteration) in the number of objects with distinct descriptions considered as vectors of attribute values, 
(iii) \textsc{Mint} is based on MDL and outputs a small set of non-redundant informative patterns.


In addition, a series of experiments shows that \textsc{Mint} is efficient and outputs sets of patterns of high quality: the patterns describe meaningful groups of objects with quite precise boundaries.
Actually, the \textsc{Mint} algorithm is able to mine numerical patterns both on small and on large datasets, and it is most of the time more efficient and reliable than its competitors \textsc{Slim} and \textsc{RealKrimp}. 
The proposed encoding scheme is based on prequential plug-in codes, which have better theoretical properties than the codes used in \textsc{Slim}, \textsc{RealKrimp} and an MDL-based method for discretization called \textsc{IPD}.


The paper has the following structure. 
In Section~\ref{sec:related_work} we discuss the state-of-the art algorithms in itemset pattern mining for numerical data.
Section~\ref{sec:basics} introduces the main notions used in the paper while in Section~\ref{sec:method} we describe the bases of proposed method. 
Next Section~\ref{sec:experiments} relates the experiments carried out for illustrating the good behavior and the strengths of \textsc{Mint}. 
Finally, Section~\ref{sec:conclusion} concludes the paper with a discussion about the potential of \textsc{Mint} and some directions for future work.


%
\section{Related Work}
\label{sec:related_work}

The problem of pattern mining has been extensively studied for binary data --itemset mining-- but remains much less explored for numerical data. 
Hence a common way to mine patterns in numerical data relies on a binarization of data and then application of itemset mining algorithms.
Meanwhile, a number of approaches was designed for mining numerical datasets possibly involving binarization and taking into account the type of the data at hand.
In this section we firstly discuss different numerical data preprocessing approaches allowing the use of itemset mining and then we discuss state-of-the art approaches in numerical pattern mining.

\subsection{Numerical Data Preprocessing}
The data preprossessing is the cornerstone for discovering patterns of good quality and relies on \textit{discretization} or \textit{binarization} tasks.


%
\paragraph{Discretization.}
Discretization relies on partitioning the range of an attribute value into intervals and then mapping the intervals into integers for preserving the order of the intervals.
The existing discretization techniques can be categorized into \textit{univariate} and \textit{multivariate} techniques.

The univariate discretization includes all the methods where attributes are considered independently. 
An attribute range may be split into intervals of equal width or equal height w.r.t. frequency.
Another way to split an attribute range is based on the \textsc{K-means} algorithm~\citep{dash2011comparative}, where some criteria is used for assessing the quality of clustering and choosing an optimal $K$. 
A more flexible approach consists in splitting based on the MDL principle~\citep{kontkanen2007mdl,rissanen1992density} which is discussed in more length below. 

The fact of considering each attribute range independently does not preserve the interaction between attributes and, as a consequence, may make some patterns not recognizable.
The multivariate discretization techniques were proposed to tackle this issue.
In~\citep{mehta2005toward,kang2006ica}, multivariate discretization is based on principal component analysis and independent component analysis, respectively. 
However, both techniques do not guarantee to take into account possible complex interactions between attributes and require some assumptions on either distribution or correlation~\citep{nguyen2014unsupervised}. 
Accordingly, authors in~\citep{nguyen2014unsupervised} propose an MDL-based algorithm for multivariate discretization which shows the shortcomings mentioned above and which works in unsupervised settings (contrasting a related approach in~\citep{fayyad93,boulle2006modl,bondu2010non}).
Indeed, MDL is used in a large number of approaches and is detailed below in~\S~\ref{subsec:mdl-pm}.

%
%

\paragraph{Binarization.}
Discretization is not the only step to accomplish before applying an itemset mining algorithm. 
Another important operation is binarization, i.e., the transformation of discrete values into binary attributes. 
Binarization should be carefully taken into account as it may affect the results of itemset mining and induce the loss of important information.
Moreover, binarization is associated with the risk of introducing artifacts and then obtaining meaningless results.

A simple and popular binarization is based on ``one-hot encoding'', where each discrete value is associated with one binary attribute. 
The number of attributes may become very large which can lead to an excessive computational load. 
Moreover, one-hot encoding does not necessarily preserve the order of discrete values.

By contrast, interordinal scaling~\citep{kaytoue2011revisiting} preserves the order of values by introducing for each discrete value $v$ two binary attributes ``$x \geq v$'' and ``$x\leq v$''. 
However, in~\citep{kaytoue2011revisiting} it was shown that, with a low-frequency threshold, mining itemsets in interordinally scaled data becomes much more expensive than mining hyper-rectangles in numerical data. Hyper-rectangles here are studied in the framework of interval pattern structures.

An alternative approach to interordinal scaling~\citep{srikant1996mining} consists in computing more general itemsets based on considered discrete values. 
The authors introduces the notion of the partial completeness w.r.t. a given frequency threshold. 
This notion allows one to formalize the information loss caused by partitioning as well as to choose an appropriate number of intervals w.r.t. chosen parameters. 
As the interordinal scaling, this approach suffers from pattern explosion. In addition, this method requires to set some parameters, e.g., frequency threshold and completeness level, however, their optimal values are unknown.

Despite its limitations, one-hot encoding remains a good option provided that suitable constraints can be defined for avoiding attribute explosion and allowing for tractable computation.

\paragraph{Numerical attribute set assessment based on ranks.} 
\begin{wrapfigure}{r}{0.1\textwidth}
  \begin{center}
    \includegraphics[width=0.1\textwidth]{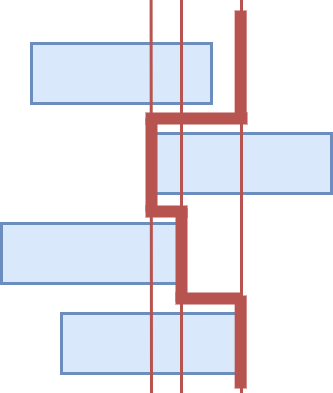}
  \end{center}
\end{wrapfigure}

One of the main drawbacks of the above  mentioned approaches that consider discretization and binarization as mandatory preprocessing steps, is that the quality of the output depends on the quality of the discretization.
In mining numerical patterns, when the boundaries of patterns are not well aligned as shown in the figure on the right, uniform boundaries will produce imprecise descriptions, while using exact boundaries may greatly complicate pattern mining.

An alternative approach that ``simulates'' multi-threshold discretization is proposed in a seminal paper~\citep{calders2006mining}.
It consists in (i) considering the ranks of attribute values instead of actual real values, and (ii) evaluating the sets of numerical attributes using rank-based measures.
In ~\citep{calders2006mining}, the authors propose several support measures based on ranks.
In such a way, the problem of dealing with concrete attribute values is circumvented by considering the coherence of the ranked attribute values.   
Moreover, in~\citep{tatti2013itemsets}, the authors propose two scores to evaluate a set of numerical attributes using ranked attribute values as well.
The scores are used to find the best combinations of attributes w.r.t. rank-based supports.

In all the methods mentioned in this subsection, patterns are understood as combinations of the attribute ranges as a whole. 
These methods do not provide descriptions related to some particular parts of the range if needed, which is the main focus of this paper.

\subsection{MDL-based approaches to Pattern Mining}
\label{subsec:mdl-pm}

The MDL principle~\citep{grunwald2007minimum} is based on the slogan: ``the best set of patterns is the set that compresses the database best''.
There is a significant amount of papers about the use of the MDL principle in pattern mining and this is very well presented in the report of Esther Galbrun \citep{Galbrun20}.
MDL has been used in many different contexts but hereafter we focus on pattern mining.
One of the most famous MDL-based itemset miners is \textsc{Krimp} which was introduced in~\citep{vreeken2011krimp}.
\textsc{Krimp} relies on two steps that consist in (i) generating a set of frequent patterns, and (ii) selecting those minimizing the total description length. 
While \textsc{Krimp} is an efficient and well-designed itemset miner, it requires that all frequent itemsets should be generated. 
Moreover, increasing the frequency threshold may lead to a worse compression, so 
the \textsc{Slim} system~\citep{smets2012slim} was proposed to tackle this issue. 
In contrasting to \textsc{Krimp}, \textsc{Slim} does not require that all itemsets should be generated in advance, since the candidates to become optimal itemsets are discovered gradually.  
Nevertheless, the encoding scheme used in \textsc{Krimp} and \textsc{Slim} shows a range of limitations that are discussed in more detail in Section~\ref{ssec:mdl}.
In continuation, the \textsc{DiffNorm} algorithm~\citep{budhathoki2015difference} is an extension of \textsc{Slim} that is based on a better encoding scheme and can be applied to a collection of datasets for finding the difference in datasets in terms of itemsets. 
Another MDL algorithm related to the \textsc{Krimp} family was proposed in~\citep{akoglu2012fast} for fixing the scalability issues.
This algorithm deals with categorical data and is less sensitive to combinatorial explosion.

All the aforementioned MDL-based algorithms represent a ``model'', i.e. a set of patterns, as a two-column table, where the left-hand column contains the pattern descriptions and the right-hand column contains the associated code words. 
Another way to store patterns is proposed in the \textsc{Pack} algorithm~\citep{tatti2008finding}, where the model is encoded as a decision tree, so that a node corresponds to an attribute.
A non-leaf node has two siblings reflecting the presence or absence of this attribute in an itemset. 
The itemset, in turn, is a path from the root to a leaf node.
One main difference between the \textsc{Pack} approach and the algorithms of the \textsc{Krimp} family is that 0's and 1's are symmetrically considered in \textsc{Pack}.


The \textsc{Stijl} algorithm~\citep{tatti2012style} is a tree-based MDL approach taking into account both 0's and 1's and storing itemsets in a tree.
However, contrasting \textsc{Pack}, \textsc{Stijl} relies on ``tiles'', i.e., rectangles in a dataset. 
The tree in \textsc{Stijl} is a hierarchy of nested tiles, where parent-child relations are inclusion relations within the set of tiles. 
A child tile is created whenever its density --the relative number of 1's-- differs a lot from the parent one.
An extension of tile discovery is proposed in~\citep{faas2020vouw} where ``geometric pattern mining'' with the \textsc{Vouw} algorithm is introduced. 
This algorithm may consider arbitrarily shaped patterns in raster-based data, i.e., data tables with a fixed order of rows and columns, and it is able to identify descriptive pattern sets even in noisy data. 
Finally, the discovery of tiles is also closely related to Boolean Matrix Factorization (BMF). 
In a nutshell, the objective of BMF is to find an approximation of a binary matrix $C$ by a Boolean product of two low-rank matrices $A$ and $B$. 
The columns of $A$ and the rows of $B$ describe the factors, which correspond to tiles. 
The MDL principle can also be applied to the BMF problem~\citep{miettinen2014mdl4bmf,makhalova2020below}.

All the MDL-based algorithms which are surveyed above apply to binary or categorical data. 
Now we focus on a few algorithms which are dealing with pattern mining in numerical data. 
First of all the \textsc{RealKrimp} algorithm~\citep{witteveen2014realkrimp} is an extension of \textsc{Krimp} to real-valued data, where patterns are based on axis-aligned hyper-rectangles. 
Even if the algorithm does not require any preprocessing, it actually needs a discretization of the data. 
Moreover, there is also a ``feature selection'' step where unimportant hyper-rectangle dimensions are removed. 
 \textsc{RealKrimp} is tailored to mine high-density patterns, and to minimize the combinatorial explosion, it constructs each hyper-rectangle using a pair of neighboring rows sampled from the original dataset. 
Then, without prior knowledge about the data, the choice of the size of sampling is difficult as a too small sample size may output very general patterns, while a too large sample size may increase the execution time. 
The problem of an inappropriate sample size may be partially solved by setting a large ``perseverance'', i.e., how many close rows should be checked to improve compression when enlarging the hyper-rectangle, and ``thoroughness'', i.e., how many consecutive compressible patterns are tolerated. 
As it can be understood, finding optimal parameters in \textsc{RealKrimp} constitutes an important problem in the pattern mining process.
Moreover, the hyper-rectangles in \textsc{RealKrimp} are evaluated independently, meaning that the algorithm searches for a set of optimal patterns instead of an optimal pattern set.
The subsequent pattern redundancy may be mitigated by sampling data and computing the hyper-rectangles in different parts of the attribute space.
Thereby, \textsc{RealKrimp} relies on many heuristics and has no means to jointly evaluate the set of generated hyper-rectangles.
In addition, heuristics imply some prior knowledge about the data which is not always available in practice.
All these aspects should be taken into account.


Another approach to mine informative hyper-rectangles in numerical data was proposed in~\citep{makhalova2019numerical}. 
The approach can be summarized in 3 steps:
(i) greedily computing dense hyper-rectangles by merging the closest neighbors and ranking them by prioritizing dense regions,
(ii) greedily optimizing an MDL-like objective to select the intervals --sides of hyper-intervals-- for each attribute independently,
(iii) constructing the patterns using the selected intervals and maximizing the number of instances described by the intervals by applying a closure operator (a closed set is maximal for a given support). 
Actually, this approach tends to optimize entropy, which is proportional to the length of data encoded by the intervals, and does not take into account the complexity of the global model, i.e., the set of patterns. 
This simplification is based on the observation that each newly added interval replaces at least one existing interval, and thus, the complexity of the model does not increase. 
Moreover, the compression process is lossy as the data values can be reconstructed only up to some selected intervals. 
Finally, the approach allows for feature selection, but does not address explicitly the problem of overlapping patterns. 

Based on this first experience, below we propose \textsc{Mint} algorithm, which is based on MDL and aimed at mining patterns in numerical data. 
We restrict the patterns to be hyper-rectangles as they are the most interpretable types of multidimensional patterns. 
As in \textsc{RealKrimp}, the \textsc{Mint} algorithm deals with discretized data which allow us to define a lossless compression scheme. 
The problem of feature selection is not addressed while patterns may overlap. 
Finally, contrasting \textsc{RealKrimp}, \textsc{Mint} is less dependent on heuristics and discovers an MDL-optimal pattern set rather than single optimal patterns. 



\section{Basics}\label{sec:basics}
\subsection{Formalization of data and patterns}\label{ssec:pattern_mining}

Let $D$ be a dataset that consists of a set of objects $G = \{g_1, \ldots, g_n\}$ and a set of attributes $M = \{m_1, \ldots, m_k\}$. 
The number of objects and attributes is equal to $n$ and $k$, respectively. 
Each attribute $m_i \in M$ is numerical and its range of values is denoted by $range(m_i)$. 
Each object $g\in G$ is \textit{described} by a tuple of attribute values $\delta(g) = \left\langle v_{i} \right\rangle_{i \in \{1, \ldots, k\}}$. 

As patterns we use  axis-aligned hyper-rectangles, or ``boxes''.
In multidimensional data, an axis-aligned hyper-rectangle has one of the simplest descriptions --a tuple of intervals--  and thus can be easily analyzed by humans. 
The hyper-rectangle describing a set of objects $B$ is given by 
$$h =  \left\langle [\min\{v_i \mid v_i \in \delta(g), g\in B \}, \max\{v_i \mid v_i \in \delta(g), g\in B \}]\right\rangle_{i \in \{1, \ldots, |M|\}}.$$

We call the $i$-th interval of a hyper-rectangle the $i$-th \textit{side} of the hyper-rectangle.
The support of a hyper-rectangle $h$ is the cardinality of the set of objects whose descriptions comply with $h$, i.e., $sup(h) = |\{ g \in G \mid \delta(g) \in h\}|$. 

Often, instead of continuous numerical data, one deals with discretized data, where the continuous range of values $range(m_i)$ is replaced by a set of integers, which are the indices of the intervals.
Formally speaking, a range $range(m_i)$ is associated with a partition based on a set of intervals  $\mathcal{B}_i = \{B_i^j = [c_{j-1}, c_{j}) \mid j = 1, \ldots, l \}$, where $c_0$ and $c_l$ are the minimum and maximum values, respectively, of $range(m_i)$. Thus, each $v \in [c_{j-1}, c_{j})$ is replaced by $j$.

The endpoints of the intervals can be chosen according to one of the methods considered above, e.g., equal-width, equal-height intervals or using the MDL principle, and the number of the intervals may vary from one attribute range to another attribute range. The endpoints make a \textit{discretization grid}. The number of the grid dimensions is equal to the number of attributes. 

A chosen discretization splits the space $\prod_{i = 1}^{|M|}range(m_i)$ into a finite number of elementary hyper-rectangles $h_e \in \{\prod_{i = 1}^{|M|} B_{i}^{j} \mid B_{i}^{j} \in \mathcal{B}_i\}$, i.e., each side of an elementary hyper-rectangle is composed of one discretization interval $B_i^j$.
Non-elementary hyper-rectangles are composed of consecutive elementary hyper-rectangles. 

For a hyper-rectangle $h = \left\langle [c^{(l)}_i, c^{(u)}_i)\right\rangle_{i \in \{1, \ldots, |M|\}}$, where $c^{(l)}_i$ and $c^{(u)}_i$ are endpoints of intervals from $\mathcal{B}_i$, in the discretized attribute space we define the \textit{size} of the $i$th side as the number of elementary hyper-rectangles included into this side, i.e., $size(h,i) = |\{B_i^j \mid B_i^j \subseteq [c^{(l)}_i, c^{(u)}_i)\}|$.
Further, we use $h$ to denote a hyper-rectangle (pattern) and $\mathcal{H}$ to denote a set of hyper-rectangles (patterns).

\begin{figure}
    \begin{minipage}[b]{0.17\linewidth}
    \begin{tabular}{cc}\toprule
        $m_1$ & $m_2$\\
        \midrule
        0.30 & 0.15\\
        0.05 & 3.90\\
        0.20 & 4.35\\
        4.40 & 0.00\\
        4.30 & 3.70\\
        4.10 & 3.90\\
        4.25 & 6.60\\
        7.10 & 4.15\\
        6.70 & 6.50\\
        6.90 & 7.40\\
        7.45 & 6.75\\
        7.10 & 7.35\\\bottomrule
    \end{tabular}
\end{minipage} 
\begin{minipage}[b]{0.25\linewidth}
    \begin{tabular}{ccc}\cmidrule[\heavyrulewidth]{1-2}
     $m_1$ & $m_2$ & \tikz [remember picture] \node (rightmark) {};\\
        \cmidrule{1-2}
        0 & 0 & \tikz [remember picture] \node (n1) {};\\*
        0 & 4 &\tikz [remember picture] \node (n2) {};\\
        0 & 4 &\tikz [remember picture] \node (n3) {};\\
        4 & 0 &\tikz [remember picture] \node (n4) {};\\\
        4 & 4 &\tikz [remember picture] \node (n5) {};\\
        4 & 4 &\tikz [remember picture] \node (n6) {};\\
        4 & 7 &\tikz [remember picture] \node (n7) {};\\
        7 & 4 &\tikz [remember picture] \node (n8) {};\\
        7 & 7 &\tikz [remember picture] \node (n9) {};\\
        7 & 7\\
        7 & 7\\
        7 & 7 &\tikz [remember picture] \node (n10) {};\\\cmidrule[\heavyrulewidth]{1-2}
    \end{tabular}
    \tikz [overlay,remember picture]\draw [decoration={brace,amplitude=1mm},decorate,thick] (n1.north -| rightmark) -- (n1.south -| rightmark) node [midway,right=2mm,align=left] {$h_1$};
    \tikz [overlay,remember picture] \draw [decoration={brace,amplitude=1mm},decorate,thick] (n2.north -| rightmark) -- (n3.south -| rightmark) node [midway,right=2mm,align=left] {$h_2$};
    \tikz [overlay,remember picture] \draw [decoration={brace,amplitude=1mm},decorate,thick] (n4.north -| rightmark) -- (n4.south -| rightmark) node [midway,right=2mm,align=left] {$h_3$};
    \tikz [overlay,remember picture] \draw [decoration={brace,amplitude=1mm},decorate,thick] (n5.north -| rightmark) -- (n6.south -| rightmark) node [midway,right=2mm,align=left] {$h_4$};
    \tikz [overlay,remember picture] \draw [decoration={brace,amplitude=1mm},decorate,thick] (n7.north -| rightmark) -- (n7.south -| rightmark) node [midway,right=2mm,align=left] {$h_5$};
    \tikz [overlay,remember picture] \draw [decoration={brace,amplitude=1mm},decorate,thick] (n8.north -| rightmark) -- (n8.south -| rightmark) node [midway,right=2mm,align=left] {$h_6$};
    \tikz [overlay,remember picture] \draw [decoration={brace,amplitude=1mm},decorate,thick] (n9.north -| rightmark) -- (n10.south -| rightmark) node [midway,right=2mm,align=left] {$h_7$};
\end{minipage} 
\begin{minipage}[b]{0.55\linewidth}
    \begin{minipage}[t]{0.49\linewidth}
        \includegraphics[width=1.03\textwidth]{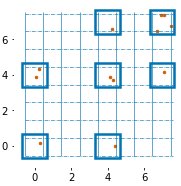}
    \end{minipage}
    \begin{minipage}[t]{0.49\linewidth}
		\centering  
		\includegraphics[width=1.\textwidth]{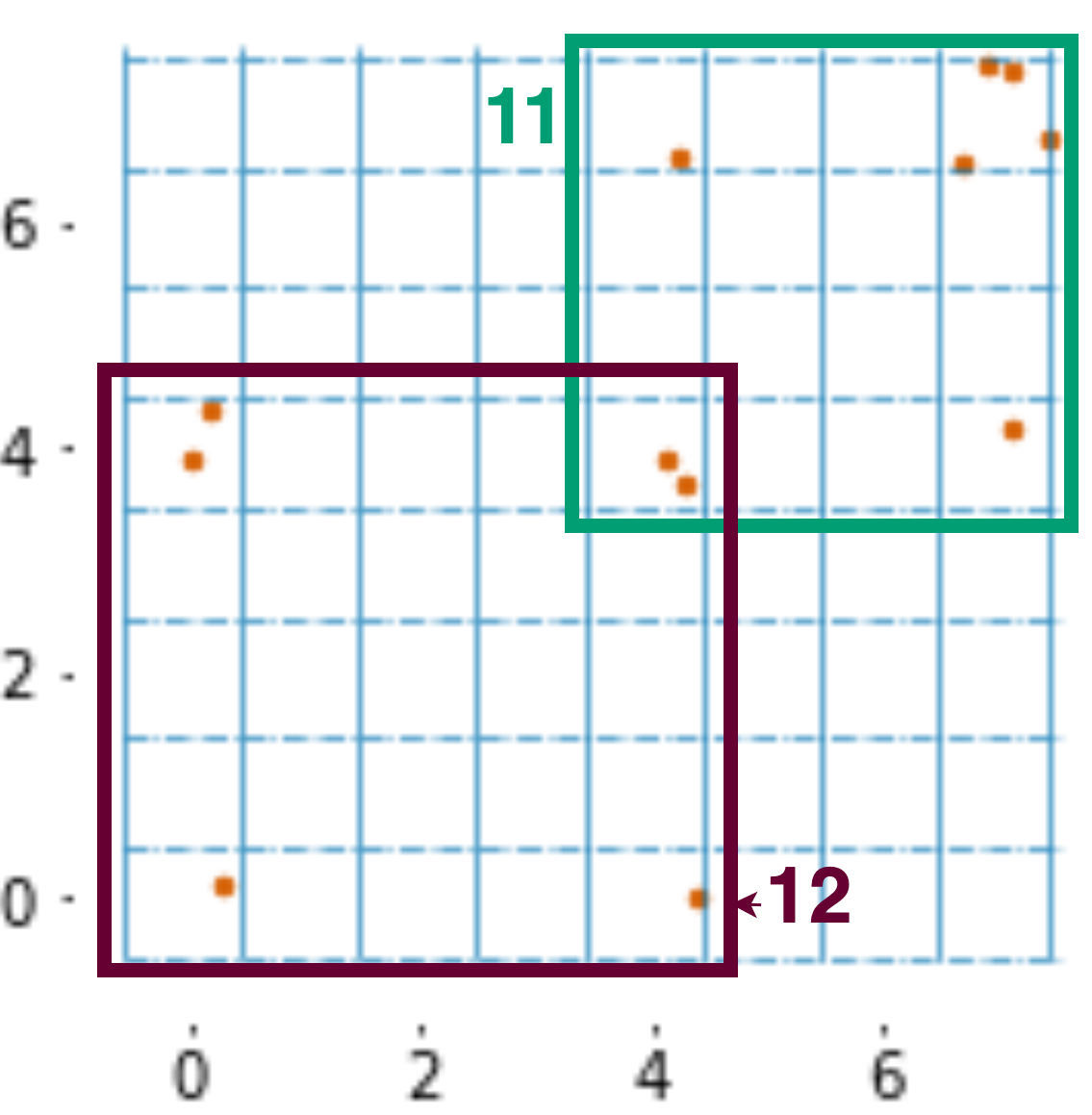}
	\end{minipage}
    \vspace{-5em}
\end{minipage}
	\caption{Dataset over attributes $\{a_1, a_2\}$ (left), its discretized version (middle), representation of the dataset on the plane and its partition into $8 \times 8$ equal-width intervals (right). The discretization grid is given by dotted lines, the corresponding non-empty elementary hyper-rectangles are given by dashed lines. The axis labels show the indices of elementary hyper-rectangles}
	\label{fig:example_notation}
\end{figure}
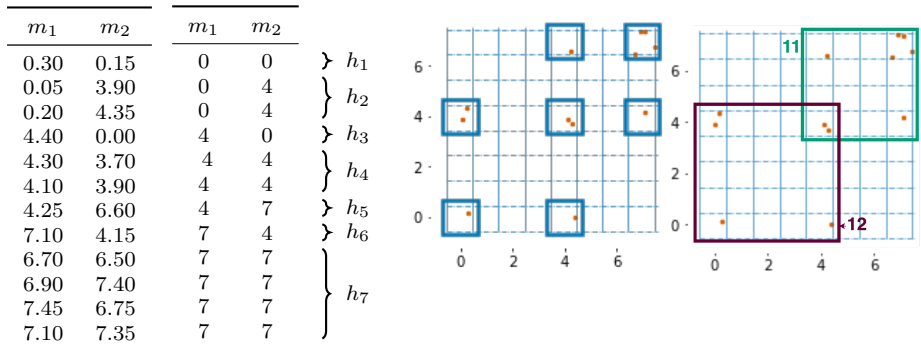

\noindent\textbf{Example}. Let us consider a dataset given in Fig.~\ref{fig:example_notation} (left). 
It consists of 12 objects described by attributes $m_1$ and $m_2$. All the descriptions are distinct (unique). 
Each attribute range is split into 8 intervals of width 1. The discretized dataset is given in Fig.~\ref{fig:example_notation} (right).
It has 7 unique rows.
The non-empty elementary hyper-rectangles correspond to non-empty squares induced by the $8 \times 8$ discretization grid.  
The number of hyper-rectangles is equal to the number of distinct rows in the discretized dataset (given in the middle).

\subsection{Information Theory and MDL}\label{ssec:mdl}

MDL~\citep{grunwald2007minimum} is a general principle that is widely used for model selection and works under the slogan ``the best model compresses data the best''. 
This principle is grounded on the following model: given a sequence that should be sent from a transmitter to a receiver, the transmitter, instead of encoding each symbol uniformly, replaces repetitive sub-sequences with code words.
Thus, instead of a symbol-wise encoded sequence, the transmitter sends a sequence of code words and a dictionary. 
The dictionary contains all used code words and the sub-sequences encoded by them. 
Using the dictionary, the receiver is able to reconstruct the original sequence. 
The MDL principle is applied to decide which sub-sequences should be replaced by the code words and which code words should be chosen for these sub-sequences.
The code words are associated in a such a way that the most frequent sub-sequences have the shortest code words.
Applied to our case, as the sequence that should be transmitted, we consider a numerical dataset, and as sub-sequences we chose patterns (hyper-rectangles).

Formally speaking, given a dataset $D$ the goal is to select such a subset of patterns $\mathcal{H}$ that minimizes the description length $L(D, \mathcal{H})$. 
In the crude version of MDL~\citep{grunwald2007minimum} the description length is given by $L(D, \mathcal{H}) = L(\mathcal{H}) + L({D} | \mathcal{H})$, where $L(\mathcal{H})$ is the description length of the model (set of patterns)  $\mathcal{H}$, in bits, and $L({D} | \mathcal{H})$ is the description length of the dataset ${D}$ encoded with this set of patterns, in bits. 

The length $L(\mathcal{H})$ characterizes the complexity of the set of patterns and penalizes high-cardinality pattern sets, while the length of data $L(D|\mathcal{H})$ characterizes the conformity of patterns w.r.t. the data. 
$L(D | \mathcal{H})$ increases when the patterns are too general and do not conform well with the data. 
Thus, taking into account both $L(\mathcal{H})$ and $L(D | \mathcal{H})$ allows to achieve an optimal balance between the pattern set complexity and its conformity with the data. 

Roughly speaking, the minimization of the total length consists in (i) choosing patterns that are specific for a given dataset, and (ii) assigning to these patterns the code words allowing for a shorter total length $L(D, \mathcal{H})$. 

In MDL, our concern is the length of the code words rather than the codes themselves. 
That is why we use a real-valued length instead of an integer-valued length.

Intuitively, the length of code words is optimal when shorter code words are assigned to more frequently used patterns. 
From the information theory, given a probability distribution over $\mathcal{H}$, the length of the Shannon prefix code for $h \in \mathcal{H}$ is given by $l(h) = -\log P(h)$ and is optimal.
Then we obtain the following probability model: given the usage $usg(h)$ of $h \in \mathcal{H}$ in the encoding. The probability distribution that ensures an optimal pattern code length for the chosen encoding scheme is given by
%
$$P(h) = \frac{usg(h)}{\sum_{h_i \in \mathcal{H}} usg(h_i)},$$
%
\noindent where the usage $usg(h)$ of a pattern $h$ is the number of times a pattern $h$ is used to cover objects $G$ in a dataset $D$. 
However, this model is based on the assumption that the total number of encoded instances (the length of the transmitting sequence) is known. 
Moreover, in order to encode/decode the message, the transmitter should know the usages $usg(h)$ of all patterns $h \in \mathcal{H}$ and the receiver should know the corresponding probability distribution, this is not usually the case.

Prequential plug-in codes~\citep{grunwald2007minimum} do not have this kind of limitation.
These codes refer to ``online'' codes since they can be used to encode sequences of arbitrary lengths and they do not require to know in advance the usage of each pattern. 
The codes are based on only previously encoded instances.
Moreover, they are asymptotically optimal even without any prior knowledge on the probabilities~\citep{grunwald2007minimum}. 
The prequential plug-in codes are widely used in recent MDL-based models~\citep{faas2020vouw,proencca2020interpretable,budhathoki2015difference}.

More formally, the idea of the prequential codes is to assess the probability of observing the $n$-th element $h^n$ of the sequence based on the previous elements $h^1, \ldots, h^{n-1}$.
Thus prequential codes allow for a predictive-sequential interpretation for arbitrary length sequences.

Let $H^n$ be the sequence $h^1, \ldots, h^{n-1}, h^{n}$. The probability of the $n$-th pattern $h^n \in \mathcal{H}$ in the pattern sequence $H^n$ is given by
\begin{equation}P_{plug\mbox{-}in}(h^n) = \frac{\prod_{h \in \mathcal{H}} [ \Gamma(usg(h) + \varepsilon) / \Gamma(\varepsilon)]}{\Gamma(usg(\mathcal{H}) + \varepsilon |\mathcal{H}|)/\Gamma(\varepsilon |\mathcal{H}|)},
    \label{eq:plug_in_final}
\end{equation}
\noindent where $usg(h)$ is the number of occurrences of pattern $h$ in the sequence $H^n$, and $usg(\mathcal{H}) = \sum_{h \in \mathcal{H}} usg(h)$ is the length of the sequence, i.e., the total number of occurrences of patterns from $\mathcal{H}$. $\Gamma$ is the gamma function. We give the technical details of the derivation of Equation~\ref{eq:plug_in_final} in Appendix~\ref{appendix:derivation_of_plugin}.

Then the length of the code word associated with $h^n$ is given as follows:
\begin{align*}
l(h^n) = -\log P_{plug\mbox{-}in}(h^n)  =\\ =\log\Gamma(usg(\mathcal{H}) + \varepsilon |\mathcal{H}|) - &\log\Gamma(\varepsilon |\mathcal{H}|) - \sum_{h \in \mathcal{H}}\left[\log\Gamma(usg(h) + \varepsilon ) - \log\Gamma(\varepsilon)\right]. \numberthis \label{eq:len_l_h}
\end{align*}
%
%
As it was mentioned above, we are interested in the \textit{length} of the code words rather than in the code words themselves. That is why we use real-valued length instead of integer-valued length for the number of bits needed to store the real code words.

To encode integers, when it is needed, we use the \textit{standard universal code for the integers}~\citep{rissanen1983universal} given by 
$L_\mathbb{N}(n) = \log n + \log \log n + \log\log \log  n + \ldots + \log c_0,$
where the summation stops at the first negative term, and $c_0 \approx 2.87$~\citep{grunwald2007minimum}. 
In this paper we write $\log$ for $\log_2$ and put $0 \log 0 = 0$.

\section{Mint}\label{sec:method}

We propose an approach to pattern mining in multidimensional numerical discretized data. 
The main assumption we rely on is that all the attributes are important, i.e., patterns are computed in the whole attribute space. 
To apply this method we consider a discretized attribute space, i.e., each attribute range is split into equal-width intervals, as it was done in~\citep{witteveen2014realkrimp,nguyen2014unsupervised}. 
The choice of the equal-width intervals is due to the fact that the cost, in bits, of the reconstruction of a real value here is constant for all intervals. 
Each object therefore is included into an $|M|$-dimensional \textit{elementary hyper-rectangle}. 
Starting from the {elementary hyper-rectangles} (each side is composed of one interval), we greedily generalize the currently best patterns and select those that provide the maximal reduction of the total description length. 
At each step we reuse some of the previously discovered candidates as well as other candidates computed on the fly using the last added pattern. 

\subsection{The model encoding }\label{ssec:approach}

Firstly, we define the total description length of the set of hyper-rectangles and the data encoded by them.
The total description length is given by $L(D, \mathcal{H}) = L(\mathcal{H}) + L(D \mid \mathcal{H})$, where $ L(\mathcal{H})$ is the description length, in bits, of the set of hyper-rectangles $\mathcal{H}$, and $L(D | \mathcal{H})$ is the description length, in bits, of the discretized dataset encoded by this set of hyper-rectangles. The initial set of the hyper-rectangles is composed exclusively of elementary ones.

To encode the set of hyper-rectangles $\mathcal{H}$, we need to encode the discretization grid and the positions of the hyper-rectangles in this grid. Thus, the total length of the pattern set is given by
\begin{align*}
L(\mathcal{H}) = & \underbrace{ L_\mathbb{N}(|M|) + \sum_{i = 1}^{|M|} L_\mathbb{N}(|\mathcal{B}_i|)}_{length\; of\; the\; grid} 
+ & \underbrace{L_\mathbb{N}(|\mathcal{H}|) + |\mathcal{H}| \left( \sum_{i = 1}^{|M|} \log \left(|\mathcal{B}_i| (|\mathcal{B}_i| + 1)/2\right) \right)}_{length\; of\; the\; pattern \; set}.
\label{eq:length_code_table}
\end{align*}
To encode the grid we need to encode the number of dimensions (attributes) $|M|$ and the number of intervals $|\mathcal{B}_i|$ within each dimension $i$. 
This grid is fixed and is not changed throughout the pattern mining process. 
To encode the pattern set $\mathcal{H}$, given the grid, we need to encode the number of patterns $|\mathcal{H}|$ and the positions of their boundaries within each dimension. Since there exist $\binom{|\mathcal{B}_i|}{2} + |\mathcal{B}_i|$ possible positions of the boundaries within the $i$-th dimension, namely $\binom{|\mathcal{B}_i|}{2}$ combinations where the boundaries are different, and $|\mathcal{B}_i|$ cases where the lower and upper boundaries belong to the same interval. 
These positions are encoded uniformly.
The latter gives the cost of $\log(|\mathcal{B}_i|(|\mathcal{B}_i| + 1))$ bits for encoding the $i$-th side of a pattern within the chosen grid.

The length of a dataset encoded with the set of patterns is given by
%
$$L(D | \mathcal{H}) =   \underbrace{L_\mathbb{N}(|G|)}_{\substack{cost\;of\;encoding\;the\\\;number\;of\;instances}}  + \underbrace{L(D(\mathcal{H}))}_{\substack{cost\;of\;encoding\\D\;with\;\mathcal{H}} } + \underbrace{L(D \ominus D(\mathcal{H}))}_{\substack{reconstruction\;cost}}$$
%
\noindent where the first component encodes the number of objects, the second one corresponds to the length of data encoded with hyper-rectangles, and the third one corresponds to the cost of the reconstruction of the object description $\delta(g) = \left\langle v_{i} \right\rangle_{i \in \{1, \ldots, |M|\}}$ up to elementary intervals. 
Let us consider the last two components in detail.

The cost of the reconstruction of the true real values is constant for all values due to the equal-width discretization and it is not changed during pattern mining, thus is not taken into account in $L(D,\mathcal{H})$. 
The dataset is encoded by exploring all objects in a given order and assigning to each object a code word of the pattern covering this object. 
According to the MDL principle, each data fragment should be covered (encoded) only once, otherwise, the encoding is redundant. 
However, some patterns might ``conflict'', i.e., include the same object.
A \textit{covering strategy} then defines which data fragment is an occurrence of which pattern. 

Here, the usage is defined as follows: $usg(h)= \{g \in G \mid g \in cover(h, G)\}$.
We discuss the covering strategy in detail in Section~\ref{ssec:algorithm}. 

From Equation~\ref{eq:len_l_h}, the length of data encoded with the plug-in codes is given by 
\begin{align*}
L(D(\mathcal{H})) = & \log \frac{\Gamma(usg(\mathcal{H}) + \varepsilon |\mathcal{H}|)/\Gamma(\varepsilon |\mathcal{H}|)}{\prod_{h \in \mathcal{H}} \Gamma(usg(h) + \varepsilon) / \Gamma(\varepsilon)} = \\
= &\log\Gamma(usg(\mathcal{H}) + \varepsilon |\mathcal{H}|) - \log\Gamma(\varepsilon |\mathcal{H}|) - \sum_{h \in \mathcal{H}} \left[\log\Gamma(usg(h) + \varepsilon) - \log\Gamma(\varepsilon)\right],
\end{align*}
\noindent where $usg(\mathcal{H}) = \sum_{h \in \mathcal{H}} usg(h)$.

Once each object has been associated with a particular pattern, its original description within the pattern up to elementary intervals is encoded in $L(D \ominus D(\mathcal{H}))$. 
We use $D \ominus D(\mathcal{H})$ to denote the difference (``distortion'') between the initially discretized dataset $D$ and the same dataset encoded with $\mathcal{H}$.

To reconstruct the dataset up to the elementary equal-width intervals we encode the positions of each object within the corresponding pattern, this cost is given by
%
    $$L(D \ominus D(\mathcal{H})) \sum_{\substack{h\in \mathcal{H}}}usg(h)\log({size(h)}) = 
    \sum_{\substack{h\in \mathcal{H}}}usg(h)\Bigg(\sum_{i = 1}^{|M|}\log({size(h,i)})\Bigg),$$%
%
\noindent where $size(h,i)$ is the number of elementary intervals that compose the side $i$ of the pattern $h$.

\noindent\textbf{Example.} Let us consider an encoding of the data by patterns according to the model introduced above for the case of the running example. 
We take the set of two hyper-rectangles $\mathcal{H} = \{h_{11}, h_{12}\}$, which are given in Fig.~\ref{fig:example_notation}. 
Let the cover of $h_{11}$ be $cover(h_{11}, G) = \{g_5, \ldots, g_{12}\}$ and cover of $h_{12}$ be $cover(h_{12}, G) = \{g_1, g_2, g_3, g_{4}\}$.
Then, the encoding of the pattern set is given by $L(\mathcal{H}) =  L_\mathbb{N}(2) + 2 \cdot L_\mathbb{N}(8) + L_\mathbb{N}(2) + 2 \cdot 2 \cdot \log 36$. 
Here, we need $2 \log 36$ bits to encode each pattern, i.e., $\log 36$ bits to encode uniformly 36 possible positions of the boundaries for each side of the hyper-rectangle.  

The length of data encoded by $\mathcal{H}$ is given by 
$L(D(\mathcal{H})) = L_{\mathbb{N}}(12) + \log \Gamma (12 + \varepsilon \cdot 2) - \log \Gamma (\varepsilon \cdot 2) - [\log \Gamma (8 + \varepsilon ) - \log \Gamma(\varepsilon) + \log \Gamma (4 + \varepsilon ) - \log \Gamma(\varepsilon)]$. 
The reconstruction error is equal to 
$L(D \ominus D(\mathcal{H})) = 8 \cdot (\log(4) + \log(4)) + 4 \cdot (\log(5) + \log(5))$, i.e., we need to encode the positions of the data points within the corresponding hyper-rectangles up to the elementary hyper-rectangles.

As we can see from the example above, the patterns can overlap. 
In such a case, one relies on a covering strategy to decide which pattern to use to encode each object.
In the next section we introduce the algorithm that defines this strategy and allows for computing patterns minimizing the total description length.

\subsection{The \textsc{Mint} Algorithm}\label{ssec:algorithm}

The objective of the \textsc{Mint} algorithm is to compute in a numerical dataset a pattern set which is the best w.r.t. the MDL principle.

\noindent \textbf{Computing Minimal Pattern Set.} Let $M$ be a set of continuous attributes, $G$ be a set of objects having a description based on attributes $M$, $\mathcal{P}$ be a set of all possible $|M|$-dimensional hyper-rectangles defined in the space $\prod_{m \in M}range(m)$, and $\emph{cover}$ be a covering strategy. 
One main problem is to find the smallest set of hyper-rectangles $\mathcal{H} \subseteq \mathcal{P}$ such that the total compressed length $L(D, \mathcal{H})$ is minimal.

The pattern search space in numerical data, where patterns are hyper-rectangles, is infinite. 
Even considering a restricted space, where all possible boundaries of the hyper-rectangles are limited to the coordinates of the objects from $G$, the search space is still exponentially large. 
The introduced total length $L(D, \mathcal{H})$ does not exhibit (anti)monotonicity property over the pattern set, and thus does not allow us to exploit some efficient approaches to its minimization. Hence, to minimize $L(D, \mathcal{H})$, we resort to heuristics.

\subsubsection{Main algorithm}

Accordingly, the main idea of \textsc{Mint} is the following.
Starting from elementary hyper-rectangles, we sequentially discover patterns that minimize the description length $L(D, \mathcal{H})$ by merging a pair of currently optimal patterns from $\mathcal{H}$.
To compute a candidate pattern based on a pair of other patterns, we introduce the \textit{join} operator $\oplus$. 
For two hyper-rectangles $h_j = \langle [ v_i^{(l)}, v_i^{(u)}) \rangle_{i \in \{1, \ldots, |M|\}}$ and  $h_k = \langle [  w_i^{(l)}, w_i^{(u)})\rangle_{i \in \{1, \ldots, |M|\}}$, their join is given by the smallest hyper-rectangle containing them, i.e., $h_j \oplus h_k = \langle [\min(v_i^{(l)}, w_i^{(l)}), \max(v_i^{(u)}, w_i^{(u)})) \rangle_{i \in \{1, \ldots, |M|\}}$.

We determine the \textit{cover} of $h_j \oplus h_k$ as the union of the covers of $h_j$ and $h_k$, i.e., $cover(h_j \oplus h_k, G) = cover(h_j, G) \cup cover(h_k, G)$.
Thus, the usage of $h_j \oplus h_k$ is simply the cardinality of its cover, i.e., $usg(h_j \oplus h_k) = |cover(h_j \oplus h_k, G)|$. Since the elementary hyper-rectangles form a partition on $G$, it follows from the definition of \textit{cover} that the usage is additive, i.e., $usg(h_j \oplus h_k) = usg(h_j) + usg(h_k)$.
Thus, each object is covered by a single pattern from $\mathcal{H}$. 
We say that pattern $h$ \textit{encodes} an object if it \textit{covers} it.

Among all candidates $\{h_j \oplus h_k \mid  h_j, h_k\in \mathcal{H} \}$ we consider as the best ones those that ensure the largest gain in the total description length:
\begin{equation}
\begin{aligned}
&\Delta L(\mathcal{H}, D, h_j, h_k) = L(D, \mathcal{H}) - L(D, \mathcal{H} \cup \{h_j \oplus h_k\} \setminus \{h_j, h_k\}) = \\
&\underbrace{L_\mathcal{N} (|\mathcal{H}|) -  L_\mathcal{N}(|\mathcal{H}| - 1) + \left( \sum_{i = 1}^{|M|} \log \left(|\mathcal{B}_i| (|\mathcal{B}_i| + 1)/2\right) \right)}_{\Delta L(\mathcal{H})} + \\
&\underbrace{\log \frac{\Gamma(|G|+ \varepsilon |\mathcal{H}|)}{\Gamma(|G|+ \varepsilon(|\mathcal{H}|-1))} + \log \frac{\Gamma( \varepsilon (|\mathcal{H}| - 1))}{\Gamma( \varepsilon|\mathcal{H}|)} - \log \frac{\Gamma (usg(h_j) + \varepsilon)\Gamma(usg(h_k) + \varepsilon)}{\Gamma(usg(h_j \oplus h_k) + \varepsilon)\Gamma(\varepsilon)}}_{\Delta L(D (\mathcal{H}))} + \\
&\underbrace{usg(h_j) size(h_j) + usg(h_k)size(h_k) - usg(h_j \oplus h_k) size(h_j \oplus h_k)}_{\Delta L(D \ominus D(\mathcal{H}))}.
\label{eq:length_gain}
\end{aligned}
\end{equation}
The term ``gain'' stands for the difference between the total description lengths obtained using the current pattern set $\mathcal{H}$ and the pattern set where patterns $h_j$ and $h_k$ are replaced by its join $h_j \oplus h_k$. The gain is the largest for the candidates that compress better a given dataset.

The minimization of $L( D | \mathcal{H})$ consists in an iterative process where candidates are computed using pairs of currently optimal patterns and selecting one that provides the largest gain in the total description length. 

The pseudocode of \textsc{Mint} is given in Algorithm~\ref{alg:mint}. 
At the beginning, the optimal patterns $\mathcal{H}$ are the elementary hyper-rectangles (line~\ref{line:init}) induced by an equal-width discretization into a chosen number of intervals $|\mathcal{B}_i|$, $i = 1, \ldots, |M|$. 
We also set an additional parameter $k$ to limit the number of candidates by $k$ nearest neighbors. 
For large datasets and large number of intervals $|\mathcal{B}_i|$, setting a low value $k << |G|$ reduces the computational efforts. 
In the discretized space, the elementary hyper-rectangles are points, thus the distance between them is the Euclidean distance. Then, given a hyper-rectangle $h_j \oplus h_k$, the neighbors are the neighbors of $h_j$ and $h_k$.
The main loop in lines~\ref{line:outer_loop}-\ref{line:end_outer_loop} consists in selecting the best candidates from the current set of candidates $\mathcal{C}$, updating the set of optimal patterns $\mathcal{H}$, and collecting new candidates in $\mathcal{C}_{new}$. 
Once all candidates from $\mathcal{C}$ have been considered in the inner loop (lines~\ref{line:inner_loop}-\ref{line:end_inner_loop}), the new candidates from $\mathcal{C}_{new}$ become current ones (line~\ref{line:change_candidates}).
In the inner loop, lines~\ref{line:inner_loop}-\ref{line:end_inner_loop}, the candidates that minimize the total description length are selected one by one. 
They are considered by decreasing gain $\Delta L$. 
At each iteration of the inner loop, the candidate $h_j \oplus h_k$  providing the largest gain $\Delta L$ is taken, the corresponding optimal patterns $h_j$ and $h_k$ are replaced with $h_j \oplus h_k$ in $\mathcal{H}$. 
The candidates based on the newly added patterns are added to $\mathcal{C}_{new}$ (line~\ref{line:new_candidates}) and are not considered at the current iteration. 
In $\mathcal{C}_{new}$ we store pairs of indices making new candidates, and only in line~\ref{line:change_candidates} we compute candidates by calculating the gains $\Delta L$ that they provide.
These gains, however, are computed only for patterns $h_j \oplus h_k$ where both $h_j, h_k$ are still present in the set $ \mathcal{H}$. 
Postponing the computation of candidates to line~\ref{line:change_candidates} allows us to reduce the number of candidates and to speed up pattern mining.
The outer loop stops when there are no more candidates in $\mathcal{C}$. After that, the pruning can be performed (line~\ref{line:mint_pruning}). We consider the details of the pruning strategy after the next example.

\begin{algorithm}[t]
	\caption{\textsc{Mint}(${D}^\ast,|\mathcal{B}_i|$, $k$, $N$)}
	\label{alg:mint}
	\begin{algorithmic}[1]
		\REQUIRE numerical dataset ${D}^\ast$, \\number of intervals for each attribute $|\mathcal{B}_i|$, $i = 1, \ldots, |M|$,\\number of the nearest neighbors for computing the candidates $k$ \\ number of candidates for pruning $N$
		\ENSURE pattern set $\mathcal{H}$,\\total description length $\mathcal{L}_{total}$
		\STATE $D \leftarrow DiscretizeData(D^\ast, \{|\mathcal{B}_i| \mid i = 1, \ldots, |M|\})$
		\STATE $\mathcal{H} \leftarrow GetElementaryHyper\mbox{-}rectangles(D)$ \label{line:init}
		\STATE $\mathcal{L}_{total} \leftarrow L(D, \mathcal{H})$
		\STATE $\mathcal{C} \leftarrow \cup_{h_j \in \mathcal{H}, h_k \in NN(h_j, k) }(h_j \oplus h_k, \Delta L(\mathcal{H}, D, h_j, h_k))$ \label{line:candidates}
		
		\WHILE{$|\mathcal{C}| > 0$} \label{line:outer_loop}
		    \STATE $\mathcal{C}_{new} \leftarrow \emptyset$
		    \STATE $(h_j \oplus h_k, \Delta L) \leftarrow PopLargestGain(\mathcal{C})$
		    \WHILE{$\Delta L > 0$} \label{line:inner_loop}
		        \IF{$h_j, h_k \in \mathcal{H}$}
		            \STATE $\mathcal{H} \leftarrow \mathcal{H} \cup \{h_j \oplus h_k\} \setminus \{h_j, h_k\}$ \label{line:change_set}
		            \STATE $\mathcal{C}_{new} \leftarrow \mathcal{C}_{new} \cup \{(h_j \oplus h_k) \oplus h \mid h \in \mathcal{H} , h \neq h_j \oplus h_k\}$\label{line:new_candidates}
		            \STATE $\mathcal{L}_{total} \leftarrow \mathcal{L}_{total} - \Delta L$
		        \ENDIF
		        \STATE $(h_j\oplus h_k, \Delta L) \leftarrow PopLargestGain(\mathcal{C})$
		  \ENDWHILE \label{line:end_inner_loop}
		  \STATE $\mathcal{C} \leftarrow \mathcal{C}_{new}$~\label{line:change_candidates}
		  \STATE \textsc{Mint-Pruning($D, \mathcal{H}, N$)} \label{line:mint_pruning}
	    \ENDWHILE \label{line:end_outer_loop}

		\RETURN $\mathcal{H}$, $\mathcal{L}_{total}$
	\end{algorithmic}
\end{algorithm}

\begin{figure}
	\centering
	\begin{minipage}[t]{0.23\linewidth}
	   \vspace{-12em}
    	 \begin{tabular}{cccc}\toprule
         \parbox[t]{2mm}{\rotatebox[origin=c]{90}{pattern}}& \multicolumn{2}{l}{\parbox[t]{2mm}{\rotatebox[origin=c]{90}{\begin{tabular}[c]{@{}l@{}}coordi-\\nates\end{tabular}}}}&\parbox[t]{2mm}{\rotatebox[origin=c]{90}{\begin{tabular}[c]{@{}l@{}}a nearest\\neighbor\end{tabular}}}\\\midrule
         $h_1$ & 0 & 0 & $h_2$ \\
         $h_2$ & 0 & 4 & $h_1$\\
         $h_3$ & 4 & 0 & $h_1$\\
         $h_4$ & 4 & 4 & $h_5$\\
         $h_5$ & 4 & 7 & $h_4$\\
         $h_6$ & 7 & 4 & $h_4$\\
         $h_7$ & 7 & 7 & $h_5$\\\bottomrule
        \end{tabular}
	\end{minipage}
	\begin{minipage}[t]{0.3\linewidth}
		\centering  
		\includegraphics[width=1.\textwidth]{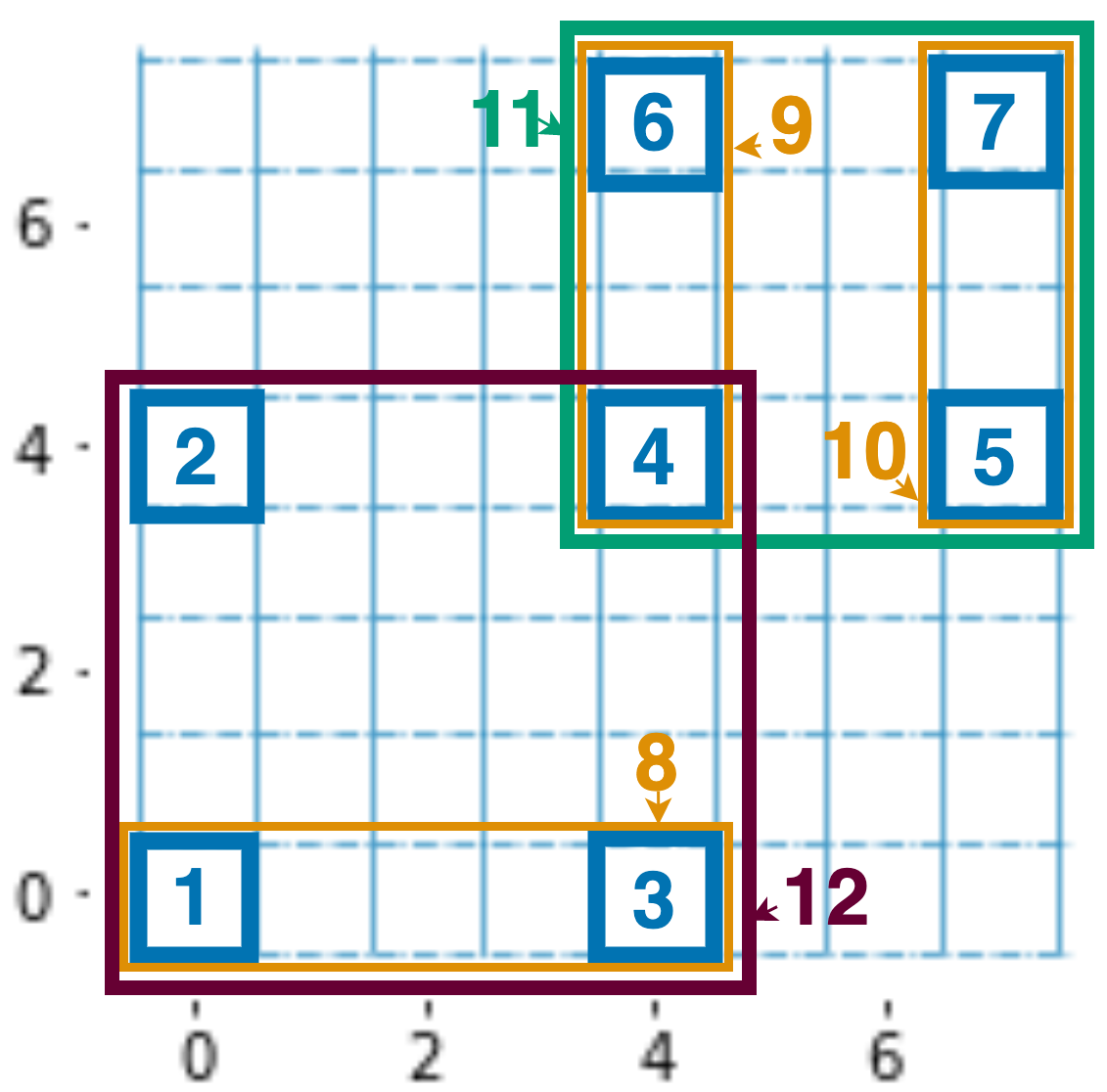}
	\end{minipage}
	\begin{minipage}[t]{0.3\linewidth}
		\centering  
		\includegraphics[width=1.\textwidth]{mdl_selected.png}
	\end{minipage}
    
    \caption{The sequential changes in the set of currently optimal patterns. The initial set of hyper-patterns is composed of elementary ones, i.e., $h_1, \ldots, h_7$}
    \label{fig:patterns_merging}
\end{figure}

\noindent\textbf{Example.} Let us consider how the algorithm works on the running example from Fig.~\ref{fig:example_notation}. 
Initially, the set of hyper-rectangles consists of elementary ones, i.e., $h_1, \ldots, h_7$ (Fig.~\ref{fig:patterns_merging}). 
We restrict the set of candidates by considering only one nearest neighbor for each pattern.
If a pattern has more than one nearest neighbor we select a pattern with the smallest index.
The nearest neighbors for each elementary hyper-rectangle are given in Fig.~\ref{fig:patterns_merging} (left).
Thus, the set of candidates is given by $\mathcal{C} = \{h_1 \oplus h_2, h_1 \oplus h_3, h_4 \oplus h_5, h_4 \oplus h_6, h_5 \oplus h_7\}$.
The patterns are added in the following order: $h_8 = h_1 \oplus h_3$, $h_9 = h_4 \oplus h_6$, $h_{10} = h_5 \oplus h_7$, which corresponds to decreasing gain.
After that, set $\mathcal{C}$ does not contain candidates that can improve the total length. 
Thus, \textsc{Mint} proceeds by considering the candidates from $\mathcal{C}_{new} = \{h_j \oplus h_k \mid j, k = 2, 8, 9, 10, j \neq k\}$, which are the candidates composed of pairs of recently added patterns and the unused old ones.
The newly added patterns are $h_{11} = h_{9} \oplus h_{10}$ and $h_{12} = h_{2} \oplus h_{8}$.
The new candidate set is $\mathcal{C}_{new} = \{h_{11} \oplus h_{12}\}$. 
The algorithm terminates,
the set of hyper-rectangles corresponding to the smallest total description length is $\mathcal{H} = \{h_{11}, h_{12}\}$.

\paragraph{Complexity of Mint.} At the beginning, the number of candidates, i.e., pairs of elementary hyper-rectangles (line~\ref{line:candidates}) is $O(min(|\mathcal{H}|^2, k \cdot |\mathcal{H}|))$, where $|\mathcal{H}|$ is the number of non-empty elementary hyper-rectangles.
The number of elementary hyper-rectangles does not exceed the number of 
objects $|G|$.
Thus, setting $k << |\mathcal{H}|$ we have the number of candidates $O(k \cdot |\mathcal{H}|) \sim (k \cdot |{G}|)$ which is linear w.r.t. the number of objects.
Further, the number of hyper-rectangles can only decrease. 
Computing a candidate $h_j \oplus h_k$ takes $O(|M|)$. 
The number of candidates added to $\mathcal{C}_{new}$ at each iteration of the inner loop is equal to $|\mathcal{H}| - 2$, $|\mathcal{H}| - 3$, etc. 
Thus, the total number of candidates is $O(|\mathcal{H}|^2)$.
The total complexity of computing candidates is $O(|M|\cdot|\mathcal{H}|^2)$. 
Searching for the maximal gain among the candidates takes $O(|\mathcal{H}|^2)$ time.
Since there can be at most $\mathcal{H}$ iterations, computing the maximal gain takes $O(|\mathcal{H}|^3)$.
In the worst case, where $|\mathcal{H}|=|G|$, the complexity is $O(|M||G|^2 + |{G}|^3)$, however it is possible either to set a small number of initial intervals $|\mathcal{B}_i|$ that ensure $|\mathcal{H}|<<|G|$ or to restrict the number of candidates. 

The theoretical complexity of the considered algorithms is estimated based on different features. In \textsc{RealKrimp}, the size of the sampling $s$ mainly affects the time complexity, the cost of computing the first hyper-rectangle is $O(|M|s^2\log s + |M||G|s)$, additional ones are mined in $O(|M|s^2 + |M||G|s)$. In the worst case, where the sample size is proportional to the number of objects, the time complexity is $O(|M||G|\log |G|)$ and  $O(|M||G|^2)$ for the first and additional hyper-rectangles, respectively.
In \textsc{Mint} the main component is the number of hyper-rectangles which is at most $|G|$, thus, the total time complexity is $O(|M||G|^2 + |G|^3)$.  Thus, both \textsc{RealKrimp} and \textsc{Slim} have polynomial complexity w.r.t. the dataset size.
\textsc{Slim} has the largest worst-case complexity $O(|C|^3 |G||I|)$, where $|C|=O(2 ^{\min(|G|, |I|)})$ is the number of the candidates, which can be exponential in the size of the dataset. Moreover, the size of the dataset used in \textsc{Slim} is larger than the size of the dataset used by \textsc{Mint} and \textsc{RealKrimp}, since the number of attributes $|I|$ in a binarized dataset is larger than the number of attributes $|M|$ in the discretized one.
Thus, \textsc{RealKrimp} and \textsc{Mint} have polynomial complexity in the input size. However, in practice, \textsc{Mint} works much faster as shown in the experiments. 

\subsubsection{Pruning strategy}

In some cases, it may happen that merging a pair of patterns does not ensure minimization of the total description length while merging several patterns at once does provide a shorter total description length. 
We propose a pruning strategy that consists in merging several candidates when the candidates generated by a pair of patterns do not provide a shorter description length.
The pseudo-code of this procedure is given in Algorithm~\ref{alg:algo}. 

The outer loop in lines~\ref{linep:outer_loop}-\ref{linep:end_outer_loop} is executed while there are some candidates allowing for a shorter total length, i.e., the number of patterns in $\mathcal{H}$ decreases. 
At each iteration of the outer loop, a new set of candidates $\mathcal{C}$ is created (line~\ref{linep:candidates}). 
As candidates we consider all pairs $h_j$, $h_k$ such that the smallest hyper-rectangle including $h_j \oplus h_k$ also contains at least one pattern $h \in \mathcal{H}$ that differs from $h_j$ and $h_k$.
We order candidates by decreasing gain $\Delta L$ and consider at the current iteration the top-$N$ candidates. 

Then, in the inner loop in lines~\ref{linep:get_top_k}-\ref{linep:end_get_top_k}, we check candidates $h_j \oplus h_k$ one by one.
In contrast to \textsc{Mint}, we do not require anymore that each candidate $h_j \oplus h_k$ improves the total description length since in lines~\ref{linep:extension}-\ref{linep:end_extension} we extend it with other patterns. 
Those patterns that improve the gain are stored in $\mathcal{S}$.
Once all hyper-rectangles contained in $h_j \oplus h_k$ have been considered, the set $\mathcal{S} \cup \{h_j \oplus h_k\}$ contains all patterns that will be replaced with $h_j \oplus h_k$, if this candidate ensures a positive gain in the total description length (lines~\ref{linep:remove_patterns}-\ref{linep:end_remove_patterns}).

\begin{algorithm}[t]
	\caption{\textsc{Mint-Pruning}(${D},\mathcal{H},N$)}
	\label{alg:algo}
	\begin{algorithmic}[1]
		\REQUIRE dataset $D$,\\pattern set $\mathcal{H}$,\\the maximal number of considered candidates $k$ 
		\ENSURE pattern set $\mathcal{H}$
		\STATE $H_{old} \leftarrow |\mathcal{H}| + 1$
		\WHILE{$H_{old} > |\mathcal{H}|$} \label{linep:outer_loop}
		    \STATE $H_{old}  \leftarrow |\mathcal{H}|$
		    \STATE $\mathcal{C} \leftarrow \{h_j \oplus h_k \mid h_j, h_k \in \mathcal{H}, \exists h \in \mathcal{H}, h \subseteq h_j \oplus h_k, h \neq h_j \neq h_k\}$\label{linep:candidates}
    		\FORALL{$h_j \oplus h_k \in GetTopNBy\Delta(\mathcal{C})$} \label{linep:get_top_k}
    		    \STATE $\mathcal{S} = \emptyset$
                \STATE $\Delta \leftarrow \Delta L(\mathcal{H}, D, h_j, h_k)$
    		    \FORALL{$h \in h_j \oplus h_k$, $h \in \mathcal{H}$} \label{linep:extension} 
    		        \STATE $cover(h_j \oplus h_k, G) = cover(h_j \oplus h_k,G) \cup cover(h, G)$
    		        \STATE $\Delta_h \leftarrow \Delta L(\mathcal{H}\setminus(\mathcal{S} \cup \{h\}), D, h_j, h_k)$
    		        \IF{$\Delta_h > \Delta$}
    		            \STATE $\mathcal{S} \leftarrow \mathcal{S} \cup \{h\}$
        	            \STATE $\Delta \leftarrow \Delta_h$
        	        \ELSE
        	            \STATE $cover(h_j \oplus h_k, G) = cover(h_j \oplus h_k, G) \setminus cover(h, G)$
        	        \ENDIF
        	   \ENDFOR\label{linep:end_extension} 
    		   \IF{$\Delta > 0$} \label{linep:remove_patterns}
    		        \STATE $\mathcal{H} = \mathcal{H} \setminus (\mathcal{S} \cup \{h_j, h_k\}) \cup \{h_j \oplus h_k\}$
    		   \ENDIF \label{linep:end_remove_patterns}
    		\ENDFOR \label{linep:end_get_top_k}
    	\ENDWHILE \label{linep:end_outer_loop}
		\RETURN $\mathcal{H}$
	\end{algorithmic}
\end{algorithm}

\paragraph{Complexity of Pruning.} At the beginning, the number of candidates is $O(|\mathcal{H}|^2)$, where $|\mathcal{H}|$ is the number of the currently optimal patterns. 
This value does not exceed the number of elementary hyper-rectangles that, in turn, is not greater than $|G|$.
We may restrict the number of considered candidates $\mathcal{C}$ to $N$ (line~\ref{linep:get_top_k}).
Thus, setting $N << |\mathcal{H}|$ we have a linear number of candidates at each iteration, however, without restrictions, the number of candidates is $O(|\mathcal{H}|^2)$. 
Computing a candidate from a pair of patterns takes $O(|M|)$. 
For each candidate we search for the included patterns, which takes $O(|M| |\mathcal{H}|)$.
Thus, in total we have $O(|\mathcal{H}|^2 (|M| + |M| |\mathcal{H}|)\sim O(|M||\mathcal{H}|^3)$ for one iteration of the inner loop. 
Since there can be at most $O(|\mathcal{H}|)$ iterations of the outer loop, the worse total complexity of pruning is $O(|M||\mathcal{H}|^4)$. However, $\mathcal{H}$ and the number of iterations of the outer loop are usually small, and pruning is performed fast.
Otherwise, the set of the candidates can be limited to the top $N$.

\section{Experiments}\label{sec:experiments}

In this section, we compare \textsc{Mint} with \textsc{Slim} and \textsc{RealKrimp} --the most similar MDL-based approaches to numerical pattern mining. 
\textsc{Slim} and \textsc{RealKrimp} are not fully comparable with our approach, since they have their own parameters that actually affect both the performance and the quality of the results.

\textsc{Slim} works with binarized data, while \textsc{Mint} and \textsc{RealKrimp} work with discretized data. 
However, \textsc{Slim} and \textsc{Mint} allow for choosing the number of discretization intervals.
\textsc{Slim} is better adapted to mine patterns in datasets with a coarse discretization. 
It should be notices that a coarse discretization results in a moderate increase in the number of attributes, while a fine discretization usually results in a drastic increase and can make the task intractable.
By contrast, \textsc{Mint} is able to mine patterns efficiently even in datasets with a fine discretization.
A last difference is that \textsc{Slim} and \textsc{Mint} evaluate a pattern set as a whole, while \textsc{RealKrimp} evaluates each pattern in a set independently. 

\paragraph{Parameters of the methods.} To compare the aforementioned methods we chose the following parameters. 
For \textsc{Slim} and \textsc{Mint}, that allow for choosing the discretization strategy, we chose the discretization into 5 and $\sqrt{|G|}$ equal-width intervals. 
The first one is expected to be more suitable for \textsc{Slim}, while the second one is expected to give better results for \textsc{Mint}. 
For \textsc{Mint} we additionally set the number of neighbors, considered for computing candidates, equal to the number of intervals.
We also use discretization by Interaction-preserving discretisation (\textsc{IPD})~\citep{nguyen2014unsupervised} --an unsupervised multivariate MDL-based discretizer-- under assumption that this method provides an optimal splitting of the attribute ranges. 
Since \textsc{Slim} deals with binary data, we transform each discretization interval into a binary attribute and use these data in \textsc{Slim}.

\textsc{RealKrimp} relies on a set of parameters, the most important among them are \textit{sample size}, \textit{perseverance}, and \textit{thoroughness}. Sample size defines the size of the dataset sample that will be used to compute patterns. We consider samples of size $\sqrt{|G|}$, 0.25$|G|$ and 0.5$|G|$. The sample size affects the running time: the smaller samples allow for faster pattern mining operations, while too small samples may prevent to discover interesting patterns. Perseverance regulates the behavior of \textsc{RealKrimp} to reach a global minimum of the description length. Large values of perseverance help to reach a global minimum. Perseverance is then set to 1/5 of the sample size. Thoroughness is the maximal number of consecutive non-compressible hyper-intervals that should be considered. We set thoroughness equal to 100. 
The parameters of the real-world datasets are given in Table~\ref{tab:datasets}. 

\begin{table}[]
    \caption{Description of datasets and the parameters of the discretization (applicable to \textsc{Mint} and \textsc{Slim}). All attributes are numerical}
    \label{tab:datasets}
    \centering
        \begin{tabular}{l|cc|ccc|c} \toprule
        \multirow{2}{*}{name} & \multirow{2}{*}{$|G|$} & \multirow{2}{*}{$|M|$} & \multicolumn{3}{c|}{\#intervals} & \multirow{2}{*}{classes} \\\cmidrule{4-6}
        &  &  & 5 & $\sqrt{|G|}$ & \textsc{IPD} &  \\\midrule
        iris & 150 & 4 & 20 & 45 & 8 & 3 \\
        wine & 178 & 13 & 65 & 163 & 26 & 3 \\
        haberman & 306 & 3 & 15 & 42 & 6 & 2 \\
        ecoli & 336 & 7 & 29 & 91 & 14 & 8 \\
        breast wisconsin & 569 & 30 & 146 & 581 & 84 & 2 \\
        spambase & 4601 & 57 & 266 & 1794 & 149 & 2 \\
        waveform & 5000 & 40 & 200 & 2681 & 198 & 3 \\
        parkinsons & 5875 & 18 & 87 & 961 & 92 & - \\
        statlog satimage & 6435 & 36 & 180 & 2335 & 213 & 6 \\
        gas sensor & 13910 & 128 & 566 & 8915 & 1060 & 27 \\
        avila & 20867 & 10 & 37 & 445 & 88 & 12 \\
        credit card & 30000 & 24 & 112 & 1401 & 195 & 2 \\
        shuttle & 58000 & 8 & 40 & 459 & 91 & 77 \\
        sensorless dd & 58509 & 48 & 219 & 5022 & 693 & 11 \\
        mini & 130064 & 50 & 117 & 1623 & 1031 & 2 \\ 
        workloads & 200000&5&25&2109&296107&-\\  \midrule
        \textbf{mean} & \textbf{33425} & \textbf{30} & \textbf{132} & \textbf{1792} & \textbf{254} & - \\\bottomrule
        \end{tabular}
\end{table}

\begin{figure}
    \centering
    \includegraphics[width=\textwidth]{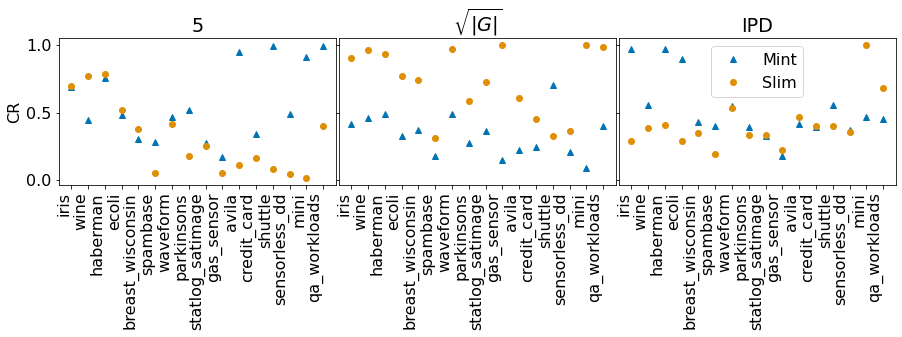}
    \caption{Compression ratios of \textsc{Slim} and \textsc{Mint} for different discretizations (into 5 intervals, $\sqrt{|G|}$ and by \textsc{IPD} discretizer), the smaller values are better}
    \label{fig:CR}
\end{figure}

\paragraph{Compression ratio.} Firstly, we consider the main quality measure of the MDL-based methods, namely \textit{compression ratio}. 
\textsc{RealKrimp} mines each pattern independently from others, thus the compression ratio of a pattern set is not measurable in this case.
\textsc{Slim} and \textsc{Mint} have different encoding schemes and compress binary and discretized datasets, respectively. Thus the compression ratios are not fully comparable. 
However, for the sake of completeness, we report the compression ratios that they provide in Fig.~\ref{fig:CR}. 
As it was expected, \textsc{Slim} works better in the case of a coarse discretization. For example, for equal-width discretization into 5 intervals, \textsc{Slim} provides much better compression than \textsc{Mint} for datasets such as ``aliva'', ``shuttle'', ``sensorless dd'', and ``mini''.
However, \textsc{Mint} works consistently better for fine discretized dataset (Fig.~\ref{fig:CR} in the middle, where the number of intervals is $\sqrt{|G|}$). 
This can be explained by the fact that \textsc{IPD} and \textsc{Mint} compress the data in the same way, while \textsc{Slim} uses an additional binarization step, thus ... 
For the \textsc{IPD}-discretized data \textsc{Slim} often ensures better compression. 
This can be explained by the fact that \textsc{IPD} compresses the same data as \textsc{Mint}, and \textsc{Mint} might be unable to compress further the same dataset. 
\textsc{Slim}, in turn, compresses the dataset that was additionally binarized.
Thus, the better compression of \textsc{Slim} may be partially explained by artifacts caused by this data transformation.

\begin{figure}
    \centering
    \includegraphics[width=\textwidth]{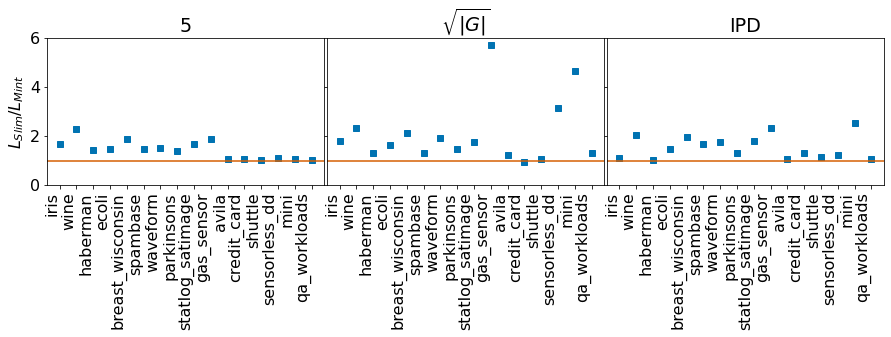}
    \caption{The ratio of the total description length of \textsc{Slim} by the total description length \textsc{Mint}. The values close to 1 (horizontal line) corresponds to the cases where the total description lengths provided by both methods are almost the same}
    \label{fig:length_gain}
\end{figure}

\paragraph{Total description length of the compressed data.} 
Another important characteristic of MDL-based approaches is the total length resulting from compression. 
Because of the different forms of data used by \textsc{Slim} and \textsc{Mint} (binary and discrete, respectively), the algorithms may have quite different compressed total description lengths.
Nevertheless, we compare them.
The ratio of the total length of \textsc{Slim} by the total length of \textsc{Mint} is given in Fig.~\ref{fig:length_gain}. 
As in the case of the compression ratio, \textsc{RealKrimp} does not allow for computing the total length of data encoded by a pattern set, it computes only length gains provided by single patterns. 
Fig.~\ref{fig:length_gain} shows that the total length of \textsc{Slim} is usually about 2 times greater than the total length of \textsc{Mint}. 
The latter can be explained, in particular, by the redundancy caused by data binarization. 
Then we may conclude that the encoding of the discretized data provided by \textsc{Mint} is more concise than the encoding of the binarized data used in \textsc{Slim}.

\paragraph{Running time.} The next important question is the running time, which is reported in Table~\ref{tab:time}. 
The cases, where pattern sets are not computed filled in the table with ``$\ldots$''. 
The performance of \textsc{Slim} and \textsc{Mint} is affected by the number of discretization intervals, while the performance of \textsc{RealKrimp} depends heavily on the sample size.

The running time reported in the table shows that for small datasets ($<1k$) all methods work fast and often complete the work in less than a second. 
For average-sized ($<50k$) and large ($>50k$) datasets the running time depends on the chosen parameters. 
\textsc{Slim} mines fast patterns in datasets with a coarse discretization (into 5 intervals). However, for fine discretizations (into $\sqrt{|G|}$ or \textsc{IPD}) the running time increases drastically. 
For example, for ``shuttle'' dataset, \textsc{Slim} terminates in 1 second, when the number of intervals is 5, while to mine patterns when the attribute ranges are split into $\sqrt{|G|}$ intervals and by \textsc{IPD}, \textsc{Slim} requires 17394 and 650 seconds, respectively.
Again for \textsc{Slim}, the scalability issues are especially pronounced for datasets with a large number of attributes, e.g., for ``gas sensor'' dataset, \textsc{Slim} terminates in 8206, \textbf{\ldots}\footnote{The experiments are not completed within one week.}, and 72488 seconds, while \textsc{Mint} requires only 43, 1088, and 1000 seconds for the same discretization parameters.

\textsc{RealKrimp} also suffers from poor scalability: for average- or large-size datasets, setting a small sample size, e.g., $\sqrt{|G|}$, does not allow to find a sufficient amount of interesting patterns, while setting a reasonable sample size ($0.25|G|$ or $0.5|G|$) results in a drastic increase of the running time and memory requirement. 
 
Our experiments show that \textsc{Mint}, dealing with the same discretization as \textsc{Slim}, requires less time to mine patterns, especially for large datasets. 
However, it is not enough to assess the performance of the patterns by considering their running time. 
It it also important to study how many patterns they return and which kind of patterns.

\begin{table}[]
\caption{Running time, in seconds, of \textsc{Slim} and \textsc{Mint} and \textsc{RealKrimp}}.
\label{tab:time}
    \centering
    \setlength{\tabcolsep}{2.5pt}
    \begin{tabular}{l|cc|rrr|rrr|rrr}\toprule
     \multirow{3}{*}{name}&\multirow{3}{*}{$|G|$}&\multirow{3}{*}{$|M|$}& \multicolumn{3}{c}{\textsc{Slim}}& \multicolumn{3}{c}{\textsc{Mint}}& \multicolumn{3}{c}{\textsc{RealKrimp}}\\\cmidrule{4-12}
     &  &  & \multicolumn{3}{c}{\# intervals}& \multicolumn{3}{c}{\# intervals}& \multicolumn{3}{c}{sample size}\\
     & & & 5 &$\sqrt{|G|}$& \textsc{IPD}& 5 &$\sqrt{|G|}$&\textsc{IPD}&$\sqrt{|G|}$&$0.25|G|$&$0.5|G|$\\\midrule
        iris & 150 & 4 & 0 & 0 & 0 & 0 & 0 & 0 & 0 & 0 & 0 \\
        wine & 178 & 13 & 1 & 0 & 0 & 0 & 0 & 0 & 0 & 0 & 0 \\
        haberman & 306 & 3 & 0 & 0 & 0 & 0 & 0 & 0 & 0 & 0 & 0 \\
        ecoli & 336 & 7 & 0 & 0 & 0 & 0 & 0 & 0 & 0 & 1 & 1 \\
        breast w. & 569 & 30 & 3 & 53 & 3 & 2 & 1 & 3 & 0 & 7 & 22 \\
        spambase & 4601 & 57 & 11 & 6159 & 63 & 6 & 88 & 117 & 6 & 14169 & 18133 \\
        waveform & 5000 & 40 & 505 & 50022 & 1346 & 5064 & 194 & 23 & 2 & 5510 & 11940 \\
        parkinsons & 5875 & 18 & 3 & 599 & 27 & 12 & 129 & 85 & 6 & 999 & 1515 \\
        statlog s. & 6435 & 36 & 790 & 32212 & 1876 & 330 & 266 & 219 & 8 & 2849 & 7140 \\
        gas sensor & 13910 & 128 & 8206 & ...$^a$& 72488 & 43 & 1088 & 1000 & 113 & 42677 & 136802 \\
        avila & 20867 & 10 & 1 & 4395 & 79 & 5 & 5796 & 1135 & 69 & 8854 & 28855 \\
        credit card & 30000 & 24 & 38 & 29422 & 17083 & 611 & 15856 & 1900 & 171 & 192057 & 410780 \\
        shuttle & 58000 & 8 & 1 & 17394 & 650 & 0 & 138 & 191 & 200 & ...$^b$& ...$^b$\\
        sensorless & 58509 & 48 & 60 & 500263 & 83819 & 52 & 137213 & 11910 & 508 & 299237 & ...$^b$\\
        mini & 130064 & 50 & 22 & ...$^a$& ...$^a$& 2 & 56006 & 18292 & 861 & ...$^b$& ...$^b$\\
        workloads & 200000 & 5 & 27 & 113195 & 8177 & 14 & 128941 & 6305 & 1031 & ...$^b$& ...$^b$\\
    \bottomrule
    \end{tabular}\\
        \begin{flushleft}{\footnotesize $^a$ Interrupted process. Experiments are not completed after one week.\\\footnotesize $^{b}$ Interrupted process because of the lack of memory (requires more than 64GB).}\\
        \end{flushleft}
\end{table}

\paragraph{Number of MDL-selected patterns.}
Intuitively, the number of patterns should be small, but sufficiently enough to describe all interesting relations between attributes. 
The numbers of MDL-selected patterns for the studied methods is reported in Table~\ref{tab:n_patterns}.
The table shows that, given the same discretization, \textsc{Slim} returns usually a larger number of patterns than \textsc{Mint}. 
As in the case of the running time, \textsc{Slim} is sensitive to a large number of attributes and in this case usually returns a much larger number of patterns than \textsc{Mint}.
For example, for ``gas sensor'' dataset \textsc{Slim} returns 1608, \textbf{\ldots}$^1$, and 9554 patterns, while \textsc{Mint}, with the same discretization settings, returns only 216, 566 and 773 patterns, respectively. 

\textsc{RealKrimp}, on the contrary, returns a much smaller number of patterns than \textsc{Mint} and \textsc{Slim}. 
For example, for ``gas sensor'' dataset it returns only 4, 30, and 49 patterns for the samples of size $\sqrt{|G|}$, $0.25|G|$, and $0.5|G|$, respectively. 
Taking into account the running time, we can conclude that with the chosen parameters, the average running time per pattern is much larger for \textsc{RealKrimp} than for \textsc{Slim} and \textsc{Mint}. 
Thus, \textsc{RealKrimp} has the highest ``cost'' in seconds of generating a pattern. 
\begin{table}[]
    \caption{The number of MDL-selected patterns by \textsc{Slim} and \textsc{Mint} for discretization into 5, $\sqrt{|G|}$ intervals and the \textsc{IPD} discretization}
    \label{tab:n_patterns}
    \centering
    \setlength{\tabcolsep}{2.5pt}
    \begin{tabular}{l|cc|ccc|ccc|ccc}\toprule
     \multirow{3}{*}{name}&\multirow{3}{*}{$|G|$}&\multirow{3}{*}{$|M|$}& \multicolumn{3}{c}{\textsc{Slim}}& \multicolumn{3}{c}{\textsc{Mint}} &\multicolumn{3}{c}{\textsc{RealKrimp}}\\\cmidrule{4-12}
     &  &  & \multicolumn{3}{c}{\# intervals}& \multicolumn{3}{c}{\# intervals}& \multicolumn{3}{c}{sample size}\\
     & &  & 5 &$\sqrt{|G|}$& IPD & 5 &$\sqrt{|G|}$&\textsc{IPD}&$\sqrt{|G|}$&$0.25|G|$&$0.5|G|$\\\midrule
        iris & 150 & 4 & 19 & 18 & 9 & 9 & 9 & 6 & 2 & 4 & 6 \\
        wine & 178 & 13 & 85 & 62 & 42 & 15 & 11 & 17 & 1 & 4 & 9 \\
        haberman & 306 & 3 & 13 & 12 & 8 & 6 & 9 & 3 & 2 & 0 & 0 \\
        ecoli & 336 & 7 & 50 & 47 & 22 & 14 & 10 & 12 & 2 & 7 & 12 \\
        breast w. & 569 & 30 & 246 & 564 & 219 & 37 & 33 & 64 & 19 & 12 & 23 \\
        spambase & 4601 & 57 & 301 & 1486 & 881 & 78 & 201 & 445 & 2 & 95 & 97 \\
        waveform & 5000 & 40 & 1726 & 4071 & 1645 & 345 & 140 & 30 & 1 & 65 & 199 \\
        parkinsons & 5875 & 18 & 256 & 1418 & 708 & 82 & 207 & 377 & 4 & 21 & 25 \\
        statlog s. & 6435 & 36 & 1322 & 6480 & 1871 & 345 & 260 & 395 & 3 & 32 & 41 \\
        gas sensor & 13910 & 128 & 1608 & ... & 9554 & 216 & 566 & 773 & 4 & 30 & 49 \\
        avila & 20867 & 10 & 135 & 2149 & 1187 & 50 & 622 & 1385 & 8 & 23 & 33 \\
        credit card & 30000 & 24 & 733 & 4002 & 3757 & 697 & 1225 & 1917 & 9 & 115 & 189 \\
        shuttle & 58000 & 8 & 57 & 2947 & 1530 & 21 & 962 & 1623 & 5 & 0 & 0 \\
        sensorless & 58509 & 48 & 571 & 14040 & 10466 & 404 & 1299 & 2257 & 4 & 27 & 0 \\
        mini & 130064 & 50 & 150 & ... & ... & 39 & 901 & 7100 & 2 & 0 & 0 \\
        workloads & 200000 & 5 & 688 & 4773 & 5889 & 297 & 3238 & 4592 & 8 & 0 & 0\\\bottomrule
    \end{tabular}\\
\end{table}
Now, let us examine the quality of the generated patterns.

\paragraph{Pattern similarity (redundancy).} 
This parameter is particularly important for \textsc{RealKrimp}, where patterns are mined w.r.t. other patterns, but evaluated independently, and there are no guarantees for avoiding selection of very similar patterns.  
To study pattern similarity, we consider the average pairwise Jaccard similarity computed w.r.t. the sets of objects that the patterns describe.
We take into account all occurrences of patterns in data rather than their usage in the data covering (which is, by definition, non-redundant).

However, the average pairwise Jaccard similarity is sensitive to the number of patterns in the set, i.e., for large pattern sets having several groups of very similar patterns, Jaccard similarity may be quite low and does not spot this ``local'' redundancy.

\begin{figure}
    \centering
    \includegraphics[width=0.9\textwidth]{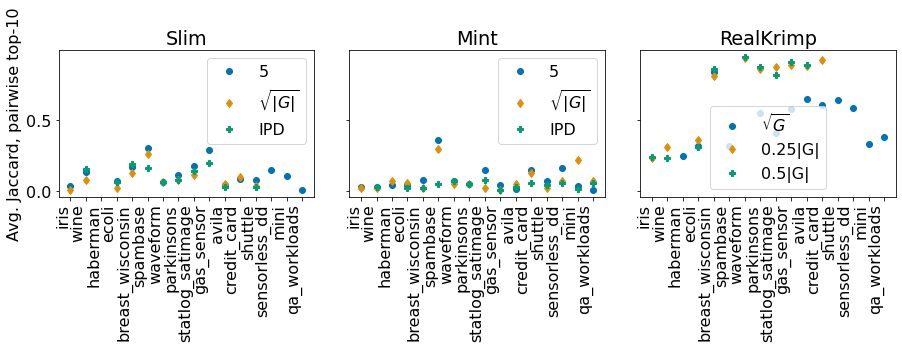}
    \caption{The average Jaccard similarity computed on the top-10 Jaccard-similar pairs for each pattern, repetitive pairs are discarded}
    \label{fig:jaccard}
\end{figure}

To tackle this issue, we do not consider pairwise similarity between all pairs, but rather the similarity between the most similar pairs, i.e. for each pattern we select at most 10 patterns among the most similar patterns w.r.t. Jaccard similarity.
Then we compute the average value of similarity by removing the repetitive pairs of patterns if they are any.
The average values of similarity are presented in Fig.~\ref{fig:jaccard}.

The results of the experiments show that on average the pairwise Jaccard similarity is the smallest for \textsc{Mint}, and only slightly higher for \textsc{Slim}. 
In \textsc{Slim} higher values are caused by the fact that each object can be covered by different non-overlapping itemsets, thus these increased values of the Jaccard similarity are partially caused by the specificity of the model.
\textsc{RealKrimp} has the largest values of the Jaccard similarity, close to 1 (see Fig.~\ref{fig:jaccard}). This result is quite expected since the patterns are evaluated independently, thus the method does not minimize redundancy in the pattern set.

\paragraph{Accuracy of patterns.} To evaluate the meaningfulness of the resulting patterns are, we measure their accuracy by considering the classes of objects they describe. 
The class labels are not used during pattern mining, and are considered only for assessing the pattern quality.

\begin{figure}
    \centering
    \includegraphics[width=0.9\textwidth]{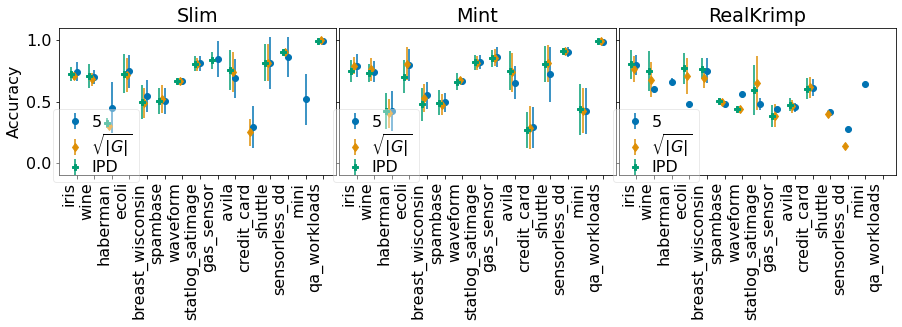}
    \caption{The average accuracy of patterns}
    \label{fig:accuracy_f1}
\end{figure}

In Fig.~\ref{fig:accuracy_f1}, the results show that \textsc{Slim} and \textsc{Mint}, being based on the same discretizations, have quite similar average accuracy.
\textsc{RealKrimp} return patterns with high accuracy for small datasets, however, loses in accuracy on large datasets.
\paragraph{Accuracy of pattern descriptions.}
As it was mentioned in the introduction, in pattern mining it is important not only to describe meaningful groups of objects, but also to provide quite precise boundaries of these groups. 
Unfortunately, we cannot evaluate how precise are the pattern boundaries for real-world dataset, since we do not have any ground truth.

\begin{figure}
    \centering
    \includegraphics[width=0.9\textwidth]{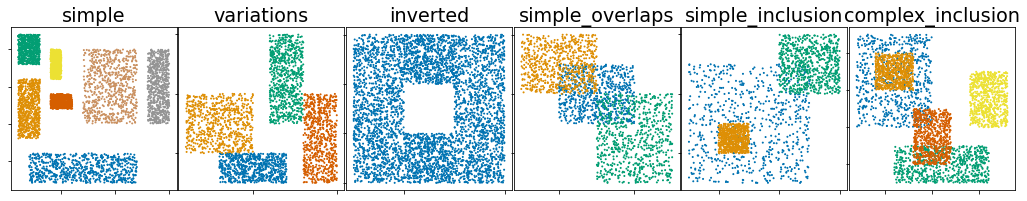}
    \caption{Six types of the generated synthetic dataset}
    \label{fig:syn_data}
\end{figure}

To evaluate how precise the boundaries of patterns we use synthetic datasets. 
We generate 6 types of 2-dimensional datasets with different number of patterns and different positions of patterns w.r.t. other patterns. 
The generated types of datasets are shown in Fig.~\ref{fig:syn_data}. The ground truth patterns are highlighted in different colors. Further we use $\mathcal{T}$ to denote a set of ground truth hyper-rectangles.
For all these types of data we generate datasets where each pattern contains 100, 200, 500, 700, 1000 objects.
In Fig.~\ref{fig:syn_data}, the ``simple'' datasets  consist of separable patterns.
The ``variations'' datasets contain adjacent patterns, and thus allow for variations in pattern boundaries. 

The ``inverted'' datasets include the most complicated patterns for \textsc{Mint} and \textsc{Slim}, since they treat asymmetrically dense and sparse regions. 
It means that these algorithms are not able to identify the hole in the middle.
Instead of this hole, we may expect a complicated description of the dense region around this hole (see Appendix~\ref{appendix:hyper_rectangle_vis}, Fig.~\ref{fig:inverted_200} for an example).
``Simple overlaps'' contains of overlapping patterns, while  ``simple inclusion'' and ``complex inclusion'' can also contain patterns which are subsets of other patterns.

For all studied heuristics, namely discretization into 5, $\sqrt{|G|}$ equal-width intervals, and an \textsc{IPD}-discretization, \textsc{Slim} returns non-empty elementary rectangles induced by the chosen discretization, i.e., it does not merge any rectangles induced by the discretization grid.
Thus, the quality of \textsc{Slim}-generated patterns is completely defined by the chosen discretization. 

For \textsc{Mint} we consider the same discretization settings as for \textsc{Slim}, namely discretization into 5, $\sqrt{|G|}$ equal-width intervals, and an \textsc{IPD}-discretization. 
We study the different settings of \textsc{Mint} in Appendix~\ref{appendix:optimal_heuristics}.
For \textsc{RealKrimp} we study only default settings: samples of size $0.5|G|$. For the discussion of the default settings of \textsc{RealKrimp} see~\citep{witteveen2012}.

We evaluate the quality of patterns using the Jaccard similarity applied to hyper-rectangles. For two hyper-rectangles $h_1$ and $h_2$ the Jaccard similarity is given by $Jaccard(h_1, h_2) = {area(h_1 \cap  h_2)}/{area(h_1 \oplus h_2)}$, where $h_1 \cap  h_2$ and $h_1 \oplus h_2$ is the intersection and join of $h_1$ and $h_2$, respectively. 


\begin{figure}
    \centering
    \includegraphics[width=1.\textwidth]{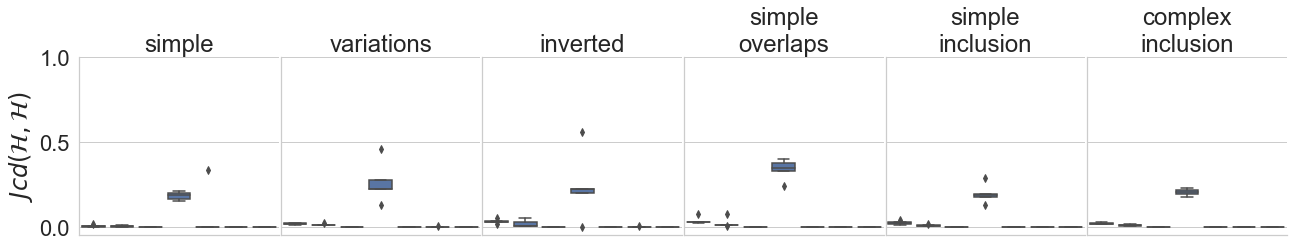}
    \includegraphics[width=1.\textwidth]{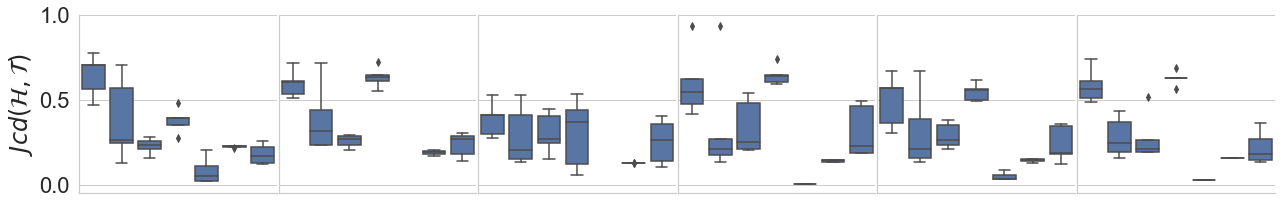}
    \includegraphics[width=1.\textwidth]{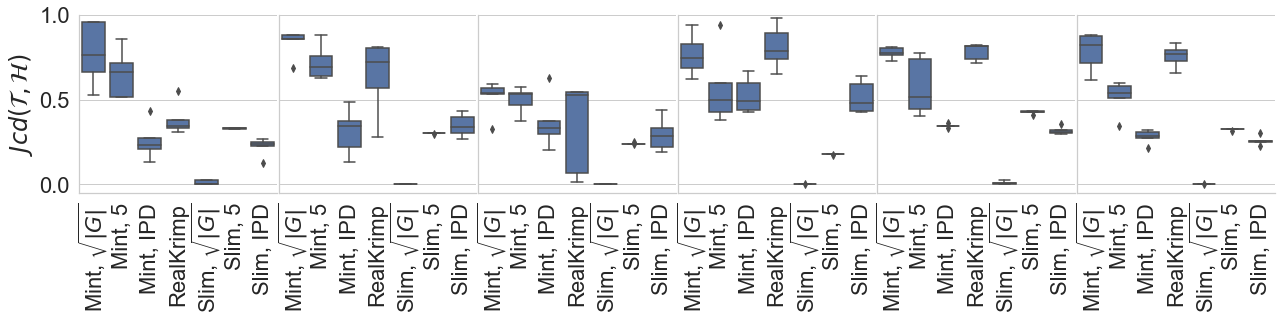}
    \caption{The average Jaccard similarity of hyper-rectangles}
    \label{fig:syn_pattern_quality}
\end{figure}

We begin with the average pairwise Jaccard similarity of the computed patterns. 
The values reported in Fig.~\ref{fig:syn_pattern_quality} (top row) show that \textsc{Slim} returns non-redundant patterns since they are non-overlapping, while \textsc{RealKrimp} returns very similar patterns.
These patterns are redundant since even for the ``simple'' datasets, where all ground truth patterns are separable and non-overlapping, the patterns returned by \textsc{RealKrimp} are very similar.
The similarity of \textsc{Mint}-selected patterns is very low, but it increases for the datasets with overlapping patterns, e.g., ``simple overlaps'' or ``simple inclusion'', and is almost 0 for the datasets with non-overlapping patterns, e.g., ``simple'' or ``variations''. 

The next question is how well the boundaries of the computed patterns from $\mathcal{H}$ are aligned with the boundaries of the ground truth patterns from $\mathcal{T}$, i.e., the patterns that we generated. 
We evaluate them by the average Jaccard similarity between the computed patterns and the most similar ground truth patterns as follows: 
\begin{equation*}
    Jcd(\mathcal{H}, \mathcal{T}) = \frac{\sum_{h_1 \in \mathcal{H}} \max_{h_2 \in \mathcal{T}}Jaccard(h_1, h_2)}{|\mathcal{H}|}
\end{equation*}.

In Fig.~\ref{fig:syn_pattern_quality} (second row) we show the average values of $Jcd(\mathcal{H}, \mathcal{T})$. 
The values close to 1 indicate that all the computed patterns are very similar to the patterns given in ground truth. 
The worst results corresponds to \textsc{Slim}: even with a ``smart'' \textsc{IPD} discretization, the boundaries computed by \textsc{Slim} are not very precise.
The low values for fine discretized data are explained by the inability of \textsc{Slim} to merge elementary hyper-rectangles.

\textsc{Mint} with the IPD-discretization returns also quite poor results. 
However, in the best settings (discretization into $\sqrt{|G|}$ intervals), \textsc{Mint} and \textsc{RealKrimp} have quite high values of $Jcd(\mathcal{H}, \mathcal{T})$.
The latter means that all patterns from $\mathcal{H}$ are quite similar to the patterns given by ground truth.

Let us study the best patterns from each set, i.e., instead of considering all patterns we evaluate the quality of the best of them.
We consider the average Jaccard similarity between the ground-truth patterns and the most similar patterns among the computed ones using the following formula:
\begin{equation*}
    Jcd( \mathcal{T}, \mathcal{H}) = \frac{\sum_{h_1 \in \mathcal{T}} \max_{h_2 \in \mathcal{H}}Jaccard(h_1, h_2)}{|\mathcal{T}|}.
\end{equation*}

The results are reported in Fig.~\ref{fig:syn_pattern_quality} (third row). The values close to 1 correspond to the case where each ground truth pattern has at least one pattern in $\mathcal{H}$ that is very similar to ground truth patterns. 
The results given in Fig.~\ref{fig:syn_pattern_quality} shows that the quality of \textsc{Mint}-generated patterns for the default discretization into $\sqrt{|G|}$ intervals is the best. 
For some datasets \textsc{RealKrimp} works equally well, e.g., ``simple overlaps'' or ``simple inclusion'', but for others it may provide quite bad results, e.g., for the simplest set of patterns contained in the ``simple'' datasets.

Comparing the values of $Jcd(\mathcal{H}, \mathcal{T})$ and $Jcd( \mathcal{T}, \mathcal{H})$ we may conclude that \textsc{RealKrimp} contains a lot of similar patterns, but these patterns do not match the ground truth patterns as well as the patterns generated by \textsc{Mint}.
Thus, the experiments shows that \textsc{Mint} (with the fine discretization) returns patterns with quite precise boundaries and outperforms the state-of-the-art MDL-based pattern miners. 

\section{Discussion and conclusion}\label{sec:conclusion}

Numerical pattern mining is a challenging task which is sometimes compared to  clustering.
However, even if the two tasks may appear as similar, in numerical pattern mining the shapes of patterns are of first importance while in clustering the main concern is the similarity between objects and the resulting groups of objects.
Accordingly, in this paper we propose a formalization of numerical pattern set mining problem based on the MDL principle and we focus on the following characteristics: (i) interpretability of patterns, (ii) precise pattern descriptions, (iii)~non-redundancy of pattern sets, and (iv) scalability.
In the paper we study and materialize these characteristics, and we also propose a working implementation within a system called \textsc{Mint}.

By ``interpretability'' we mean not only the ability to explain why a particular pattern is selected, but also the ease of analyzing a set of discovered numerical patterns for a human agent.  
With this regard, patterns of arbitrary shapes (e.g., \cite{faas2020vouw}) or even polygons (e.g.,~\citep{belfodil2017mining}) may be not an appropriate choice when considering multidimensional numerical data.
This is why we decided to work with one of the most common shapes, namely ``hyper-rectangles'', which are currently used in numerical pattern mining and related tasks~\citep{witteveen2014realkrimp,kaytoue2011revisiting}.

Another important requirement is that the boundaries of patterns should be ``well-defined'' and ``quite precise''.
A common approach to numerical pattern mining consists in firstly a data binarization and secondly a reduction to itemset mining.
Such an approach suffers from various drawbacks among which (i) the boundaries of patterns are not well-defined and this heavily affects the output, (ii) the scalability is not good because of the potential exponential number of attributes due to scaling, (iii) the information loss related to the loss of the interval order within a range may be very important\ldots
In our experiments we compare the behavior of \textsc{Mint} with the MDL-based itemset set miner \textsc{Slim} (associated with a scaling of numerical data).
The experiments demonstrate that \textsc{Slim} generally provides quite poor patterns.
Actually, when the discretization is too fine, \textsc{Slim} is not able to merge patterns into large patterns, while when the discretization is too coarse the algorithm returns very imprecise boundaries.
In addition, we also consider another MDL-based algorithm, namely \textsc{RealKrimp}, which is, to the best of our knowledge, the only MDL-based approach dealing with numerical pattern mining without any prior data transformation.
However, one main drawback of \textsc{RealKrimp} is that it mines patterns individually, and then the resulting patterns are very redundant.

Furthermore, in the experiments, both \textsc{RealKrimp} and \textsc{Slim} show a poor scalability.
\textsc{Mint} may also have a high running time for some large datasets, but still staying at a reasonable level.

\textsc{Mint} may appear to be similar to \textsc{IPD} --for ``Interaction-Preserving Dis\-cretiza\-tion''-- but both systems perform different tasks.
\textsc{Mint} could work in collaboration with \textsc{IPD} since the latter does not return exactly patterns but mainly MDL-selected boundaries.
The elementary hyper-intervals induced from \textsc{IPD} results are only fragments of ground truth patterns.
Then \textsc{Mint} could be applied to merge these elementary hyper-rectangles into larger hyper-rectangles.

Indeed, our experiments show that the data compressed by \textsc{IPD} can be even more compressed in applying \textsc{Mint}, i.e., the patterns as computed by \textsc{IPD} should still be completed for being comparable to those discovered by \textsc{Mint}.
However, as the experiments show it, directly applying \textsc{Mint} to fine discretized data allows to obtain better results than applying \textsc{IPD} as a preprocessing step.
This can be explained by the fact that \textsc{IPD} returns uniform or global boundaries, which are less precise than the boundaries specifically ``tuned'' by \textsc{Mint} for each pattern.

For summarizing, the \textsc{Mint} algorithm shows various very good capabilities w.r.t. its competitors, among which a good behavior on fine discretized datasets, a good scalability, and it outputs a moderate number of non-redundant patterns with precise boundaries.
However, there is still room for improving \textsc{Mint}, for example in avoiding redundant patterns and in the capability of mining sparse regions in the same manner as dense ones.

Future work may be followed in several directions.
Here, \textsc{Mint} works with an encoding based on prequential plug-in codes.
It could be interesting to reuse other another encoding and to check how the performance of the system evolve, trying to measure what is the influence of the encoding choice.
Moreover, we should consider more datasets and especially large and complex datasets, and try to measure the limit of the applicability of \textsc{Mint}, for in turn improving the algorithm in the details.
In general, more experiments should still be considered for improving the quality the of \textsc{Mint} algorithm.
Another interesting future direction is to use \textsc{Mint} in conjunction with clustering algorithm.
This could be a good way of associating descriptions or patterns with the classes of individuals that are discovered by a clustering process.
In this way a description in terms of attribute and ranges of values could be attached to the discovered clusters and complete the definition of the set of individuals which are covered.
This could be reused in ontology engineering for example, and as well in numerous tasks where clustering is heavily used at the moment.

\bibliography{refs}

\begin{appendices}

\section{Derivation of the plug-in codes}\label{appendix:derivation_of_plugin}

Let $H^n$ be the sequence $h^1, \ldots, h^{n-1}, h^{n}$. 
The idea of the prequential codes is to assess  the probability of observing the $n$-th element $h^n$ of the sequence $H^n$ based on the previous elements $h^1, \ldots, h^{n-1}$:
   $ P_{plug\mbox{-}in}(h^n) = \prod_{i=1}^{n}P(h^i | h ^{i-1}).$



Initially, a uniform distribution over $\mathcal{H}$ is defined with a pseudo-count $\varepsilon$ over the set of patterns $\mathcal{H}$, i.e., the probability of $h^0$ is given by $P(h^0) = \frac{\varepsilon}{\varepsilon \cdot |\mathcal{H}|}$. 
Then, during the process of transmitting/receiving messages the pattern probabilities and lengths are updated w.r.t. the patterns observed so far. 

At each single step, the distribution $P$ over $\mathcal{H}$ is multinomial with parameters $(\theta_{1}, \ldots, \theta_{|\mathcal{H}|})$, where $\theta_{i}$ corresponds to a pattern $h_i \in \mathcal{H}$. 
With the lower indices we arbitrarily enumerate patterns in $\mathcal{H}$.
The upper indices denote the sequence number of patterns in the transmitting sequence. 
Further, we will see that the order of patterns in the sequence $h^1, \ldots, h^n $ does not affect the length of the encoded sequence. This length depends only on the number of times each pattern appears in the sequence.

Taking into account the initial probabilities, the maximum-likelihood estimates of the parameters of the multinomial distribution, given the sequence $H^n = h^1\ldots h^n$, are the following:
\begin{equation}
    \hat\theta_{h^n}(h) = \frac{usg(h |h^1 \ldots h^n) + \varepsilon}{\sum_{h^\ast \in \mathcal{H}}usg(h^\ast| h^1 \ldots h^n) + \varepsilon |\mathcal{H}|},
    \label{eq:param_theta}
\end{equation}
\noindent where $usg(h | h^1 \ldots h^n)$ is the number of occurrences of pattern $h$ in the sequences observed so far (up to the $n$-th pattern inclusive).
The ML estimates from Equation~\ref{eq:param_theta} are equivalent to the probability estimates of a pattern $h$ based on its frequency in $H^h$ with Laplace smoothing having parameter $\varepsilon$~\citep{manning2008introduction}. 

Thus, taking as the probability model the multinomial distribution with the parameters estimated according to the maximum likelihood principle, the plug-in probability of the $n$-th pattern in the sequence $H^n$ is given by 
\begin{equation}
    P_{plug\mbox{-}in}(h^n) =  \prod_{i = 1}^{n}P(h^i |h ^{i-1}) = \prod_{i = 1}^{n}P_{\hat{\theta}_{h^{i-1}}}(h^i).
    \label{eq:plug_in}
\end{equation}
Combining together Equations~\ref{eq:param_theta} and~\ref{eq:plug_in} we obtain the following probability of the $n$-th pattern $h^n \in \mathcal{H}$ in the pattern sequence $H^n$:
\begin{equation}P_{plug\mbox{-}in}(h^n) = \frac{\prod_{h \in \mathcal{H}} \prod_{j = 0}^{usg(h) - 1}(j + \varepsilon)}{\prod_{j = 0}^{usg(\mathcal{H}) - 1}(j + \varepsilon |\mathcal{H}|)} = \frac{\prod_{h \in \mathcal{H}} \Gamma(usg(h) + \varepsilon) / \Gamma(\varepsilon)}{\Gamma(usg(\mathcal{H}) + \varepsilon |\mathcal{H}|)/\Gamma(\varepsilon |\mathcal{H}|)},
    \label{eq:plug_in_final1}
\end{equation}
\noindent where $usg(h)$ is the number of patterns in $H^n$, and $usg(\mathcal{H}) = \sum_{h \in \mathcal{H}} usg(h)$ is the length of the sequence, i.e., the total number of occurrences of patterns from $\mathcal{H}$. 
Then, the code length of $h^n$ is given as follows:
\begin{align*}
l(h^n) = -\log P_{plug\mbox{-}in}(h^n) = &\sum_{i = 1}^{n} - log P_{\hat{\theta}_{H^{i-1}}}(h^i) =\\ =\log\Gamma(usg(\mathcal{H}) + \varepsilon |\mathcal{H}|) - &\log\Gamma(\varepsilon |\mathcal{H}|) - \sum_{h \in \mathcal{H}}\left[\log\Gamma(usg(h) + \varepsilon ) - \log\Gamma(\varepsilon)\right].
\end{align*}

The length $l(h^n)$ can be interpreted as the sum of the log loss of the prediction errors made by sequentially predicting $h^{i}$ based on the predictor in the family of multinomial distributions over $\mathcal{H}$ that would have been the best for sequentially predicting the previous patterns $h^1, \ldots, h^{i-1}$.

\section{The ML estimates of the parameters of the multinomial distribution}\label{appendix:ml_estimates}

Let $\mathcal{H}$ be a set of $n$ patterns, i.e., $n = |\mathcal{H}|$ and $x_i$ be the number of times $h_i \in  \mathcal{H}$ appears in the sequence of patterns of length $N$. Then, the probability mass function of the multinomial distribution for the given sequence has the following form: 
$$p(x_1, \ldots, x_n \mid p_1, \ldots, p_n) = \frac{N!}{x_1! \ldots x_n!} \prod_{i = 1}^{n}p_i ^{x_i}.$$
The log-likelihood function then is the following:
$$ \ell(p_1, \ldots, p_n) = \log(N!) - \sum_{i = 1}^{n}\log (x_i!) + \sum_{i = 1}^{n} x_i \log (p_i).$$
To find the ML estimates we need to maximize this function w.r.t. $p_1, \ldots, p_n$ given the constraint on the probabilities $\sum_{i = 1}^{n}i = 1$. Applying the methods of Lagrange multipliers we get the following equation:
$$ \mathcal{L}(p_1, \ldots, p_n, \lambda) = \ell(p_1\ldots, p_n) + \lambda (1 -  \sum_{i = 1}^{n} (p_i)).$$  
Thus 
\begin{equation*}
    \begin{cases} 
    \frac{\partial \mathcal{L}}{\partial p_i} = \frac{x_i}{p_i} - \lambda = 0, i = 1, \ldots, n\\
    \frac{\partial \mathcal{L}}{\partial \lambda} = 1 - \sum_{i = 1}^{n}p_i = 0
    \end{cases}    
\end{equation*}
Taking into account the pseudo-counts, we get $x_i = usg(h_i) + \varepsilon$, where $usg(h_i)$ is the number of times the pattern $h_i$ appears in the sequence. 

Thus, 
\begin{equation*}
    \begin{cases} 
    p_i = \frac{usg(h_i) + \varepsilon}{\lambda}, i = 1, \ldots, n\\
    \sum_{i = 1}^{n}p_i = 1
    \end{cases}
\end{equation*}

Putting the last equation into the first one we get $p_i = \frac{usg(h_i) + \varepsilon}{\sum_{i = 1}^{n}{usg(h_i}) + \varepsilon  |\mathcal{H}|}$.

\section{Optimal number of intervals and neighbors}\label{appendix:optimal_heuristics}

In \textsc{Mint} we use two parameters that can affect the results: the number of discretization intervals and the number of neighbors that are used to compute the candidates.  
For a dataset of $|G|$ objects and $|M|$ attributes, we consider the following numbers of dicretization intervals: 5, $0.5 \sqrt{|G|}$, $\sqrt{|G|}$, $2 \sqrt{|G|}$, and the following numbers of nearest neighbors: 5, $0.5 \sqrt{|G|}$, $\sqrt{|G|}$, $2 \sqrt{|G|}$, $0.5 \sqrt{|G||M|}$, $\sqrt{|G||M|}$, $2 \sqrt{|G||M|}$.

\begin{figure}[h!]
    \centering
    \begin{minipage}[b]{0.32\textwidth}
        \includegraphics[width=1.\textwidth]{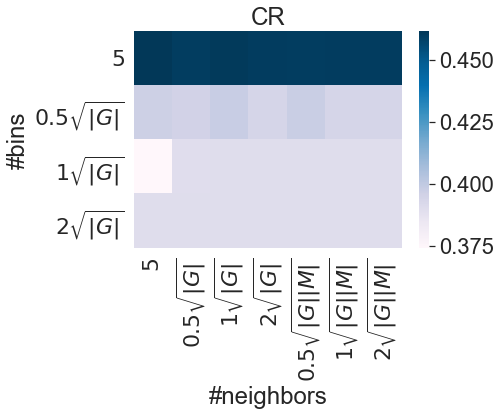}
    \end{minipage}
    \begin{minipage}[b]{0.32\textwidth}
        \includegraphics[width=1.\textwidth]{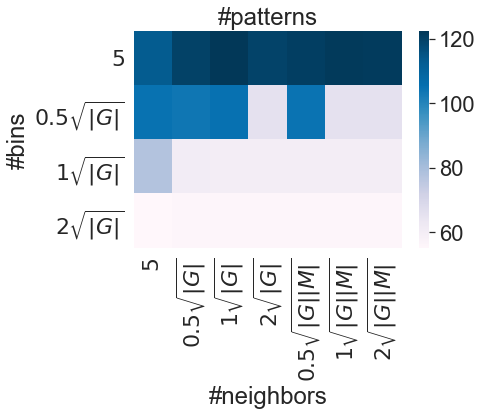}
    \end{minipage}
    \begin{minipage}[b]{0.32\textwidth}
        \includegraphics[width=1.\textwidth]{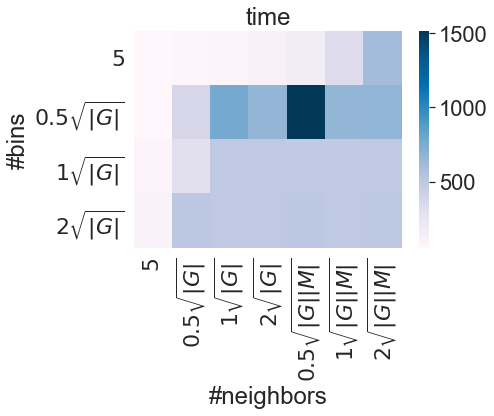}
    \end{minipage}
    \begin{minipage}[b]{0.32\textwidth}
        \includegraphics[width=1.\textwidth]{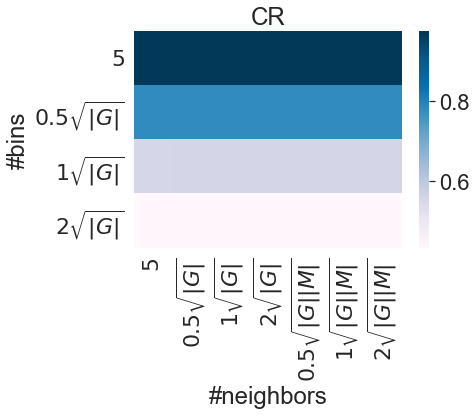}
    \end{minipage}
    \begin{minipage}[b]{0.32\textwidth}
        \includegraphics[width=1.\textwidth]{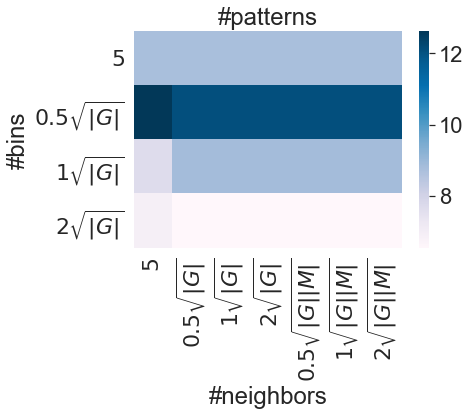}
    \end{minipage}
    \begin{minipage}[b]{0.32\textwidth}
        \includegraphics[width=1.\textwidth]{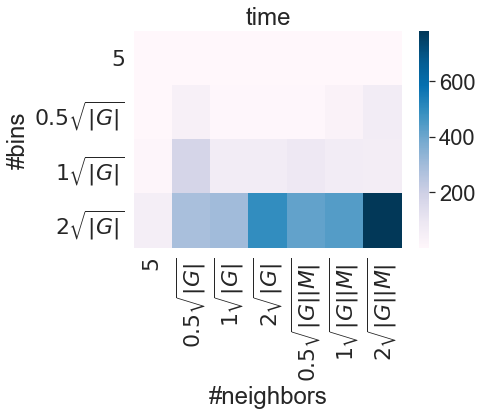}
    \end{minipage}
    \caption{Average compression ratio, number of patterns and running time of real-world (top) and synthetic (bottom) dataset for different combinations of the parameters of \textsc{Mint}.}
    \label{fig:sh}
    \centering
    \begin{minipage}[b]{0.32\textwidth}
        \includegraphics[width=1.\textwidth]{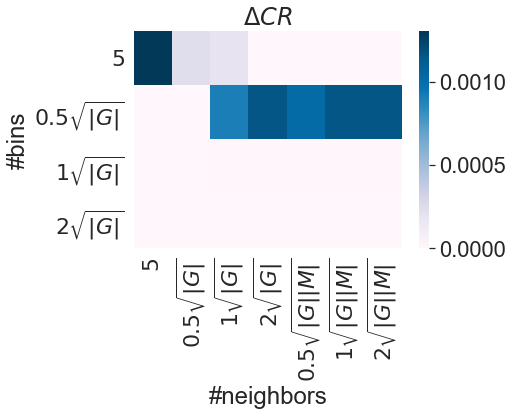}
    \end{minipage}
    \begin{minipage}[b]{0.32\textwidth}
        \includegraphics[width=1.\textwidth]{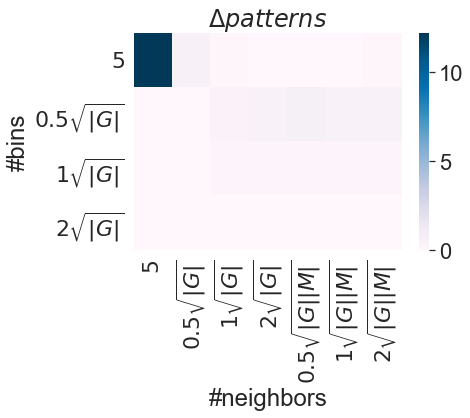}
    \end{minipage}
    \begin{minipage}[b]{0.32\textwidth}
        \includegraphics[width=1.\textwidth]{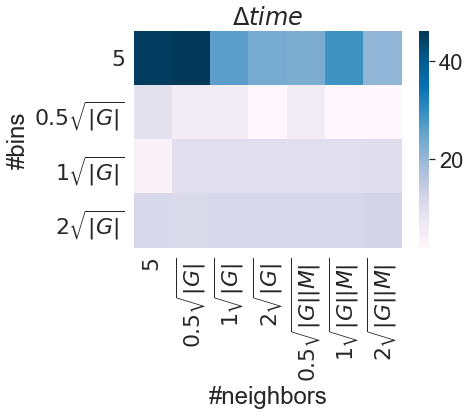}
    \end{minipage}
    \centering
    \begin{minipage}[b]{0.32\textwidth}
        \includegraphics[width=1.\textwidth]{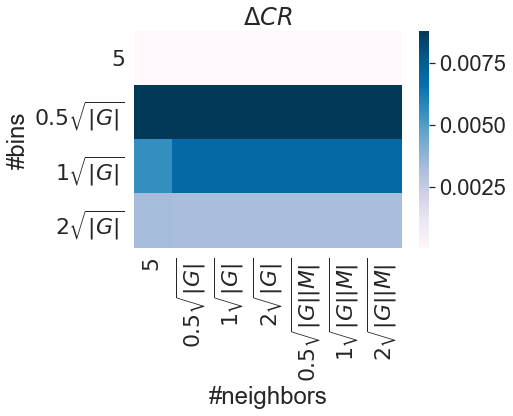}
    \end{minipage}
    \begin{minipage}[b]{0.32\textwidth}
        \includegraphics[width=1.\textwidth]{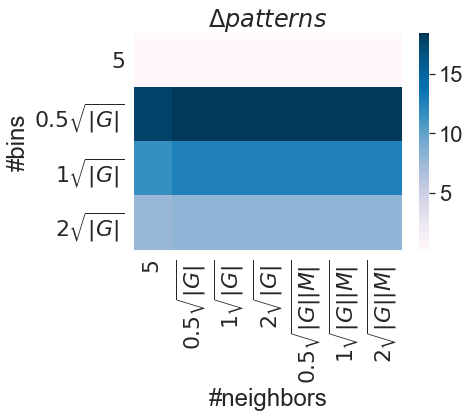}
    \end{minipage}
    \begin{minipage}[b]{0.32\textwidth}
        \includegraphics[width=1.\textwidth]{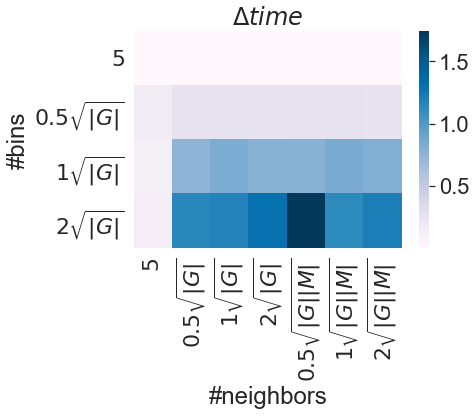}
    \end{minipage}
    \caption{Average gains in compression ratio and number of patterns for real-world (top) and synthetic (bottom) datasets achieved by applying \textsc{Mint-Pruning}. The running time of \textsc{Mint-Pruning} is reported on the right figure}
    \label{fig:sh_d}
\end{figure}

We evaluate the performance of \textsc{Mint} w.r.t. the aforementioned heuristics by the following characteristics: compression ratio, the number of patterns and the running time. The averages values of 9 real-world datasets (``iris'', ``waveform'', ``wine'', ``breast wisconsin'', ``spambase'', ``parkinsons'', ``haberman'', ``ecoli'', ``statlog satimage'') are given in Fig.~\ref{fig:sh}.
We also considered 6 types synthetic datasets (see Fig.~\ref{fig:syn_data}), where each pattern has support 100, 200, 500, 700, and 1000 (i.e., 30 synthetic datasets).

The results given in Fig.~\ref{fig:sh} show that a very coarse discretization is less suitable for \textsc{Mint}, since the compression ratio and the number of patterns is the largest (worst). 

For fine discretized datasets the number of neighbors does no not affect a lot the results. 
It appears that with a large number of discrezited intervals, the results of \textsc{Mint} is less affected by the chosen heuristics.
Taking into account the running time, the best (and the most ``stable'') results are achieved by splitting the attribute range into $\sqrt{|G|}$ intervals and by taking $\sqrt{|G|}$ nearest neighbors to compute pattern candidates.

We study how useful is the pruning strategy for reducing the compression ratio and the number of patterns, and measure how much time it takes.
The results presented in Fig.~\ref{fig:sh_d} show that the pruning strategy is useful for mining hyper-rectangles in datasets with a coarse discretization.
For fine discretized datasets, the compression ratio and the number of patterns reduce only slightly.

\begin{figure}[t!]
    \centering
    \includegraphics[width= 1.\textwidth]{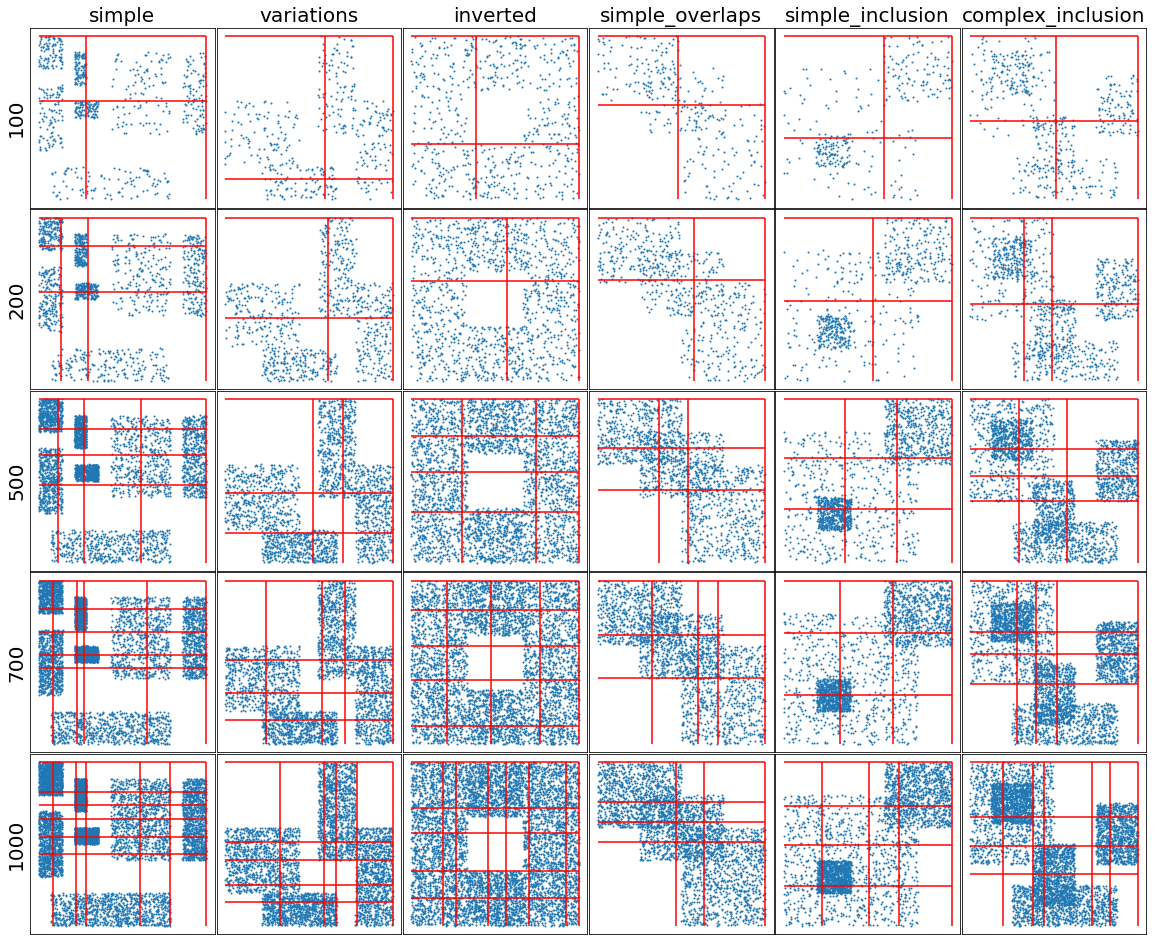}
    \caption{The results of IPD discretization for the synthetic data}
    \label{fig:IDP}
\end{figure}

\section{Visualization of some hyper-rectangles for synthetic data}\label{appendix:hyper_rectangle_vis}

In this section we give some examples of the hyper-rectangles discovered by \textsc{Slim}, \textsc{Mint}, and \textsc{RealKrimp}.
As it is mentioned above, uniform boundaries used for an itemset miner such as \textsc{Slim}) may provide poor results, i.e., the boundaries of the hyper-rectangles can be very imprecise.
And this is especially the case for the \textsc{IPD} discretization.
Actually \textsc{IPD} returns MDL-selected boundaries which are selected more for the overall distribution of points rather than for specific patterns.
The results of \textsc{IPD} are given in Fig.~\ref{fig:IDP}.
The figure shows that \textsc{IPD}, especially for sparse data, provides quite imprecise boundaries.
But even for dense data, the boundaries are not very well aligned to the ground truth patterns.
Then we cannot expect good results for \textsc{Slim} and \textsc{Mint} when they are applied to \textsc{IPD}-discretized data.

\begin{figure}
    \centering
        \includegraphics[width=1.\textwidth]{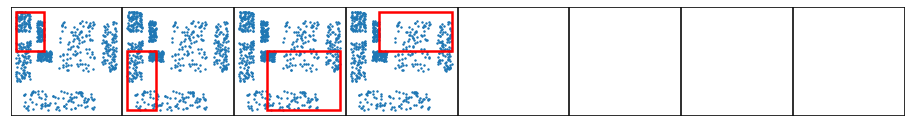}
        \caption{The patterns selected by \textsc{Mint} and \textsc{Slim} (they are the same) for \textsc{IPD}-discretized dataset ``Simple'' where support of each ground truth pattern is equal to 100}
    \label{fig:IPD_100}
\end{figure}

\begin{figure}
    \centering
        \includegraphics[width=1.\textwidth]{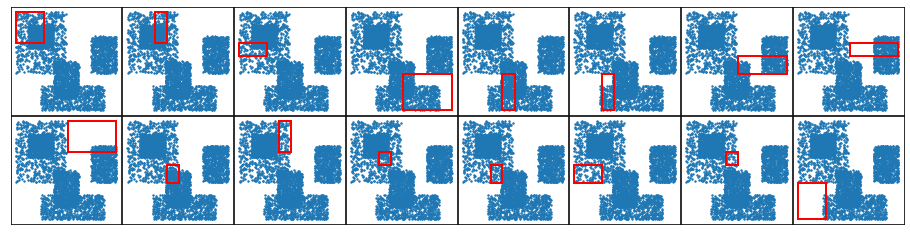}
    \caption{The patterns selected by \textsc{Mint} and \textsc{Slim} (they are the same) for \textsc{IPD}-discretized dataset ``Complex inclusion'' where support of each ground truth pattern is equal to 700}
    \label{fig:IPD_700}
\end{figure}

Let us consider the patterns computed by \textsc{Slim} and \textsc{Mint} based on IPD-discretized data. Our experiments show that for the synthetic data \textsc{Slim} and \textsc{Mint} return quite similar patterns (see Fig.~\ref{fig:IPD_100} and~\ref{fig:IPD_700}). We do not consider the \textsc{IPD}-discretized datasets further, since for the remaining datasets the anticipated results can be derived from the discretization grid given in Fig.~\ref{fig:IDP}.

\begin{figure}
    \centering
    \includegraphics[width= 1.\textwidth]{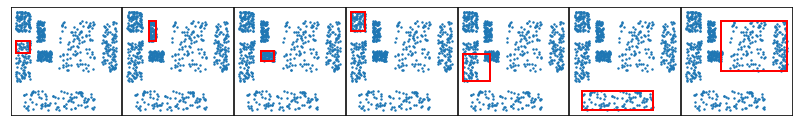} (a) \textsc{Mint} ($\sqrt{|G|}$)\\
    
    \includegraphics[width= 1.\textwidth]{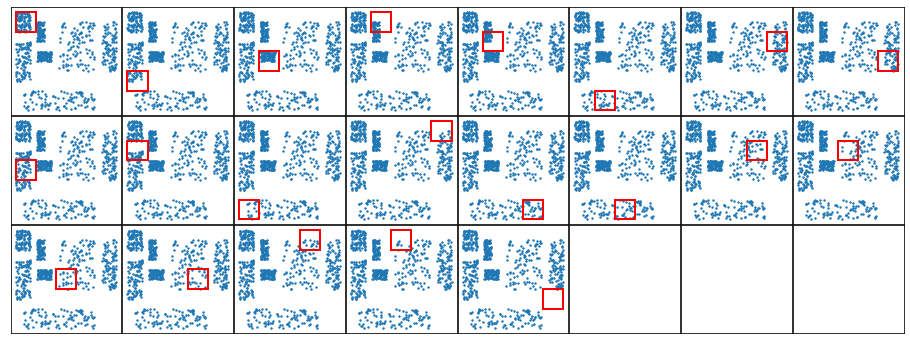} (b) \textsc{Slim} (5)\\
    
    \includegraphics[width= 1.\textwidth]{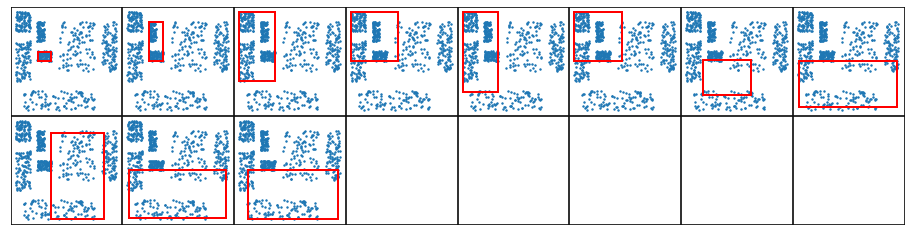} \\(c) \textsc{RealKrimp} (sampling of size $0.5 |G|$)
    \caption{The results of pattern mining for ``Simple'' dataset, support of the ground truth patterns is 100}
    \label{fig:simple_100}
\end{figure}

\begin{figure}
    \centering
    \includegraphics[width= 1.\textwidth]{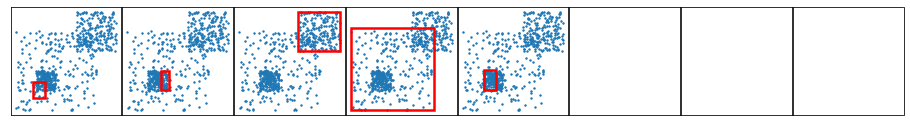}\\ (a) \textsc{Mint} ($\sqrt{|G|}$)\\
    
    \includegraphics[width= 1.\textwidth]{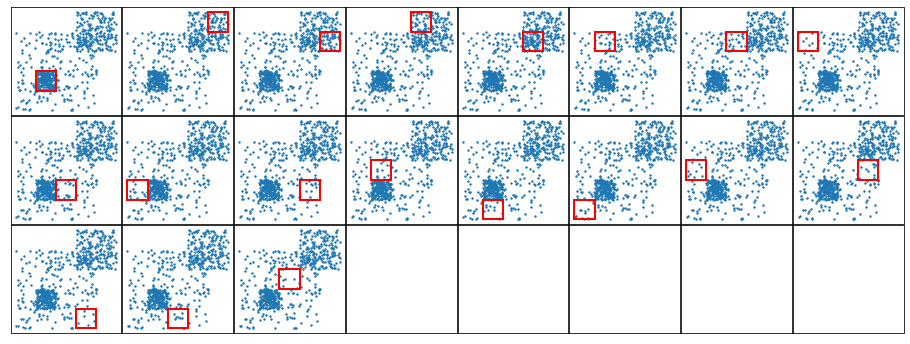} (b) \textsc{Slim} (5)\\
    
    \includegraphics[width= 1.\textwidth]{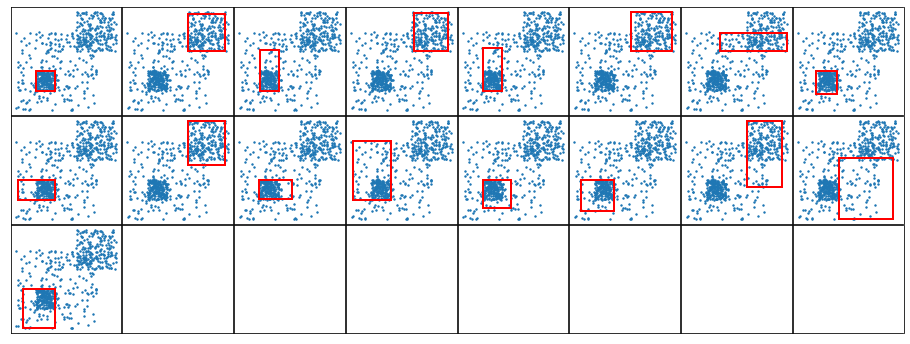} (c) \textsc{RealKrimp} (sampling of size $0.5 |G|$)

    \caption{The results of pattern mining for ``Simple inclusion'' dataset, support of the ground truth patterns is 200}
    \label{fig:simple_inclusion_200}
\end{figure}

\begin{figure}
    \centering
    \includegraphics[width= 1.\textwidth]{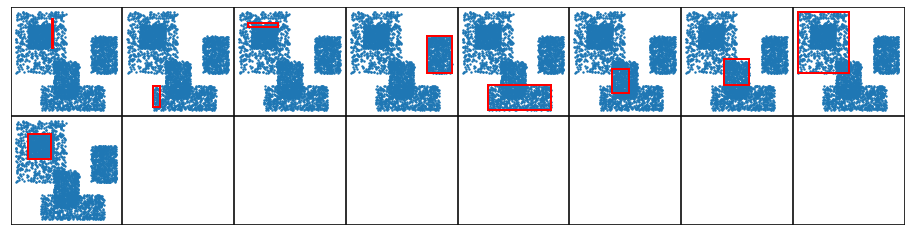} (a) \textsc{Mint} ($\sqrt{|G|}$)\\
    \includegraphics[width= 1.\textwidth]{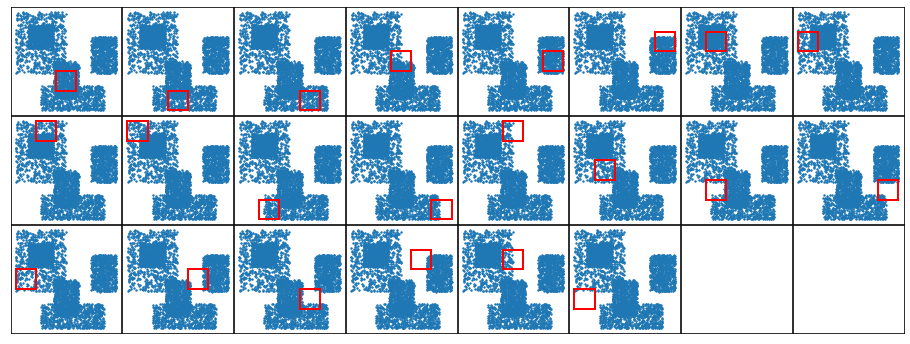} (b) \textsc{Slim} (5)\\
    \includegraphics[width= 1.\textwidth]{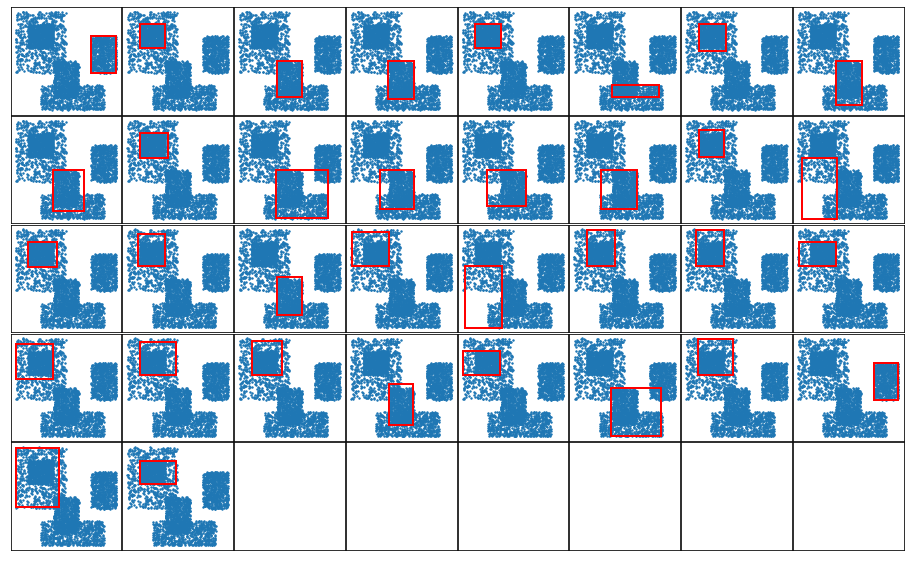} (c) \textsc{RealKrimp} (sampling of size $0.5 |G|$)
    \caption{The results of pattern mining for ``Complex inclusion'' dataset, support of the ground truth patterns is 700}
    \label{fig:complex_inclusion}
\end{figure}

\begin{figure}
    \centering
    \includegraphics[width= 1.\textwidth]{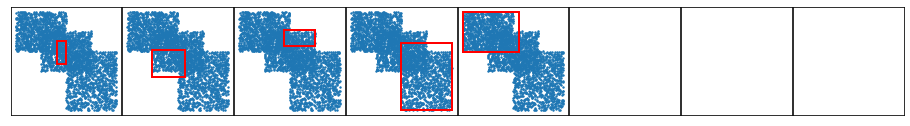}  (a) \textsc{Mint} ($\sqrt{|G|}$)\\
    \includegraphics[width= 1.\textwidth]{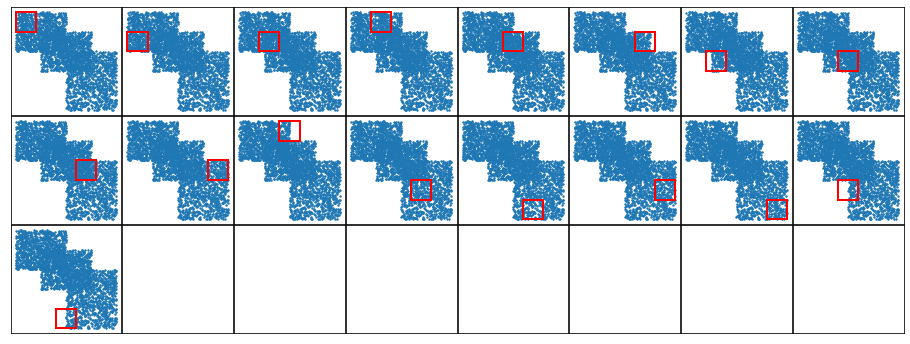} (b) \textsc{Slim} (5)\\
    \includegraphics[width= 1.\textwidth]{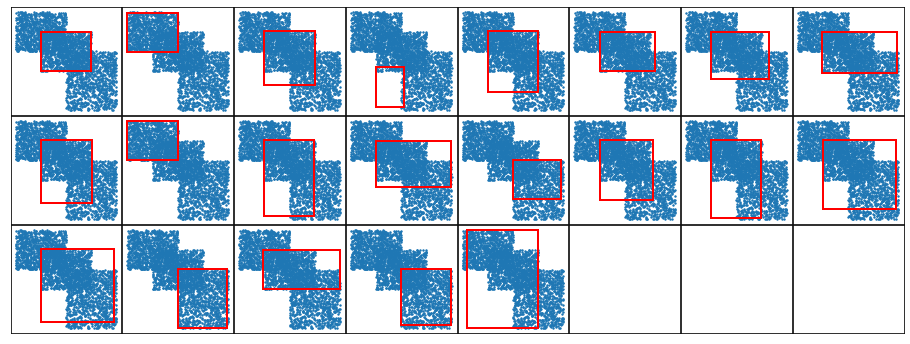} (c) \textsc{RealKrimp} (sampling of size $0.5 |G|$)
    \caption{The results of pattern mining for ``Simple overlaps'' dataset, support of the ground truth patterns is 1000}
    \label{fig:simple_overlaps_1000}
\end{figure}

\begin{figure}
    \centering
    \includegraphics[width= 1.\textwidth]{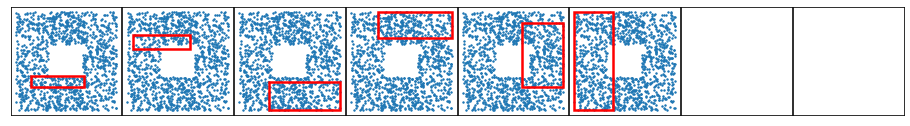} (a) \textsc{Mint} ($\sqrt{|G|}$)\\
    
    \includegraphics[width= 1.\textwidth]{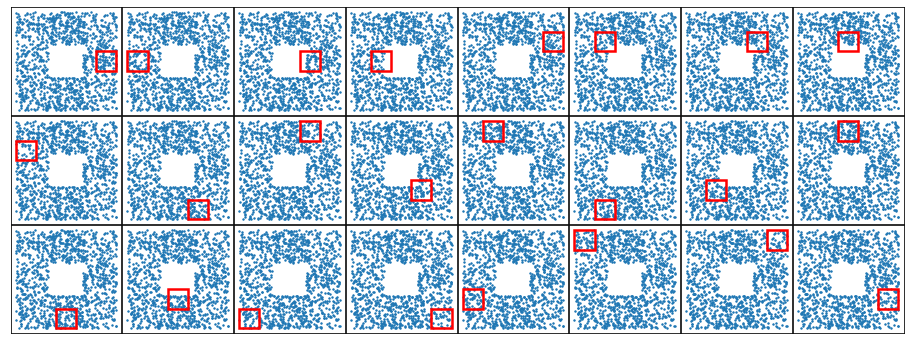} (b) \textsc{Slim} (5)\\
    
    \includegraphics[width= 1.\textwidth]{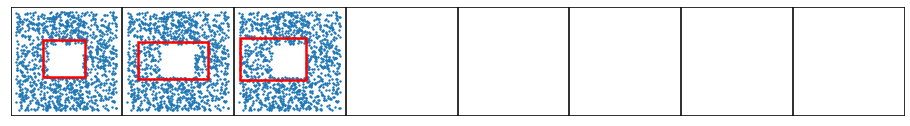} \\(c) \textsc{RealKrimp} (sampling of size $0.5 |G|$)
    \caption{The results of pattern mining for ``Inverted'' dataset, the number of data points is 200}
    \label{fig:inverted_200}
\end{figure}

Due to limited space, we show patterns for randomly chosen datasets in Fig.~\ref{fig:simple_100}-\ref{fig:inverted_200}. The ground truth patterns are highlighted in different colors in Fig.~\ref{fig:syn_data}.
As we can see \textsc{Slim} returns usually very imprecise patterns. This is a typical limitation of itemset mining methods applied to numerical data. 

\textsc{RealKrimp} returns patterns with much precise boundaries, however a lot of patterns are almost the same. For example, for ``Simple'' dataset, Fig.~\ref{fig:simple_100}, some patterns generated by \textsc{RealKrimp} are very redundant and imprecise. The patterns returned by \textsc{Mint} are much more precise. They are non-redundant, however there are patterns with imprecise boundaries (the 5th and the last one). The 1st pattern describes only a fragment of the ground truth pattern.

For dataset ``Simple inclusion'', \textsc{Mint} returns two patterns that correspond exactly to the ground truth patterns and the remaining three patterns describe the third pattern. Nevertheless, their union gives a quite correct pattern. \textsc{RealKrimp} also distinguishes only 2 patterns correctly. However, it returns a lot of similar patterns. It is important to notice that for this pattern set it is hard to find a combination of patterns that allows for reconstructing the third pattern. Thus, in the case of \textsc{RealKrimp}, redundant patterns pose a greater problem than in the case of \textsc{Mint}.

The conclusion similar to the previous one, can be drawn for Fig.~\ref{fig:complex_inclusion} and~\ref{fig:simple_overlaps_1000}. Both \textsc{Mint} and \textsc{RealKrimp} return patterns aligned to the ground truth patterns, however, the pattern sets returned by \textsc{Mint} are much less redundant, while \textsc{RealKrimp} returns very similar patterns such that without knowing the ground truth it might be hard to choose those that are the closest to the correct ones.

The last case that we consider is given in Fig.~\ref{fig:inverted_200}. It contains a sparse pattern that \textsc{Mint} and \textsc{Slim} are unable to describe, while \textsc{RealKrimp} can easily do that. As we can see from the figure, instead of a sparse hyper-rectangle, \textsc{Mint} returns a cover of the dense region around this pattern. It is the limitation of \textsc{Mint}. 
\textsc{RealKrimp} return a simple description --the exact sparse pattern-- as well as two noisy patterns.

\end{appendices}
\end{document}